\documentclass[journal,onecolumn]{IEEEtran}
\usepackage{amsthm}
\usepackage{amsmath}
\usepackage{mathrsfs}
\usepackage{amssymb} 
\usepackage{bbm} 
\usepackage{multirow} 
\usepackage{multicol}
\usepackage{mathtools, cuted}
\usepackage{units}
\usepackage{stmaryrd}   
\usepackage{verbatim}
\usepackage{enumerate} 
\usepackage[table]{xcolor}
\usepackage{slashbox} 
\usepackage{ wasysym }
\usepackage{tikz}
\usetikzlibrary{calc}
\usepackage{tabularx}

\newcommand\blfootnote[1]{%
  \begingroup
  \renewcommand\thefootnote{}\footnote{#1}%
  \addtocounter{footnote}{-1}%
  \endgroup
}

\makeatletter
\newcounter{IEEE@bibentries}
\renewcommand\IEEEtriggeratref[1]{%
  \renewbibmacro{finentry}{%
    \stepcounter{IEEE@bibentries}%
    \ifthenelse{\equal{\value{IEEE@bibentries}}{#1}}
    {\finentry\@IEEEtriggercmd}
    {\finentry}%
  }%
}
\makeatother

\usepackage{xcolor}
\usepackage{hyperref} 
\hypersetup{
    colorlinks,
    linkcolor={blue!80!black},
    citecolor={green!50!black},
    urlcolor={blue!80!black}
}

\usepackage{graphicx}
\graphicspath{{images/}}

\title{The Arbitrarily Varying Broadcast Channel 
 with Degraded Message Sets with Causal Side Information at the Encoder}
\author{\IEEEauthorblockN{Uzi Pereg and Yossef Steinberg}
}

\usepackage{biblatex}
\bibliography{IEEEabrv,references2}
\renewbibmacro{in:}{} 										

\definecolor{light-gray}{gray}{0.8}
\definecolor{dark-gray}{gray}{0.3}
\usepackage{accents}
\newlength{\dhatheight}

\newcommand{\prob}[1]{\Pr\left(#1\right)}
\newcommand{\given}{\mid}
\newcommand{\cprob}[2]{\Pr\left(#1\given #2\right)}
\newcommand{\E}{\mathbb{E}}
\newcommand{\eps}{\varepsilon}

\newcommand{\ie}{\emph{i.e.} }
\newcommand{\eg}{\emph{e.g.} }
\newcommand{\etal}{\emph{et al.} }
\newcommand{\cf}{\emph{cf.} }

\newcommand{\tm}{\widetilde{m}}	
\newcommand{\tM}{\widetilde{M}}																				

\newcommand{\tp}{\widetilde{p}}

\newcommand{\tY}{\widetilde{Y}}

\newcommand{\tU}{\widetilde{U}}

\newcommand{\tf}{\widetilde{f}}
\newcommand{\tg}{\widetilde{g}}
\newcommand{\tfnu}{\tf^{\nu}}
\newcommand{\gnu}{\tg}
\newcommand{\bR}{\tR}

\newcommand{\tR}{\widetilde{R}}

\newcommand{\hm}{\hat{m}}
\newcommand{\hq}{\widehat{q}}

\newcommand{\hgamma}{\widehat{\gamma}}

\newcommand{\hP}{\hat{P}}
\newcommand{\hM}{\hat{M}}

\newcommand{\Aset}{\mathcal{A}}
\newcommand{\Bset}{\mathcal{B}}

\newcommand{\Dset}{\mathcal{D}}

\newcommand{\Uset}{\mathcal{U}}

\newcommand{\Qset}{\mathcal{Q}}

\newcommand{\Sset}{\mathcal{S}}
\newcommand{\Wset}{\mathcal{W}}
\newcommand{\Xset}{\mathcal{X}}
\newcommand{\Yset}{\mathcal{Y}}

\newcommand{\Eset}{\mathcal{E}}
\newcommand{\markovC}[1]{%
\begin{tikzpicture}[#1]%
\draw (0,0.3ex) -- (1ex,0.3ex);%
\draw (0.5ex,0.3ex) circle (0.2ex);
\draw[white] (0.2ex,0) -- (0.5ex,0);%
\end{tikzpicture}%
}
\newcommand{\Cbar}{\markovC{scale=2}}

\newcommand{\interior}[1]{\text{int}\hspace{-0.01cm}\big( #1 \big)}

\theoremstyle{remark}	\newtheorem{theorem}{Theorem}
\theoremstyle{remark}	\newtheorem{lemma}[theorem]{Lemma}
\theoremstyle{remark}	\newtheorem{coro}[theorem]{Corollary}
\theoremstyle{remark} \newtheorem{definition}{Definition}
\theoremstyle{remark} 
\theoremstyle{remark} \newtheorem{example}{Example}

\newcommand{\avc}{\Wset}																		
\newcommand{\opC}{\mathbb{C}}																
\newcommand{\inC}{\mathsf{C}}															 	
\newcommand{\inR}{\mathsf{R}}

\newcommand{\pSpace}{\mathcal{P}}														

\newcommand{\dK}{k}

\newcommand{\enc}{f}																				
\newcommand{\dec}{g}																			 	
\newcommand{\code}{\mathscr{C}}															
\newcommand{\gcode}{\mathscr{C}^{\,\Gamma}}									
\newcommand{\tcode}{\widetilde{\code}}											

\newcommand{\cerr}{P_{e|s^n}^{(n)}}													
\newcommand{\err}{P_e^{(n)}}															
\newcommand{\tset}{\Aset^{\delta}}													
\newcommand{\qn}{q}
\newcommand{\tQ}{\hat{\Qset}_n}														

\newcommand{\enci}{f_i}																			
\newcommand{\encn}{f^n}																			
\newcommand{\encs}{\xi}																			

\newcommand{\emp}{\hP}																		  

\newcommand{\rstarC}{																			  
\, \hspace{-0.3cm} \text{ $$ \mbox{  
\hspace{-0.1cm} 
\small $\star$   
} $$ }
\hspace{-0.25cm}}

\newcommand{\sdelta}{\delta_q^{\text{\tiny $(1)$}}}
\newcommand{\wdelta}{\delta_q^{\text{\tiny $(2)$}}}

\newcommand{\sCondQ}{\mathscr{T}^{\Qset}}
\newcommand{\sCond}{\mathscr{T}}

\newcommand{\bc}{W_{Y_1,Y_2|X,S}}
\newcommand{\sbc}{W_{Y_1|X,S}}
\newcommand{\wbc}{W_{Y_2|X,S}}
\newcommand{\nBC}{W_{Y_1^n,Y_2^n|X^n,S^n}} 
\newcommand{\tBset}{\Bset}
\newcommand{\Brp}{\tBset^q} 
\newcommand{\Brpnc}{\tBset^q_{n-c}} 
\newcommand{\Bcompound}{\tBset^\Qset} 
\newcommand{\BcompoundP}{\tBset^{\pSpace(\Sset)}} 
\newcommand{\Bcompoundig}{\Bcompound_0}												
\newcommand{\avbc}{\tBset}																		
\newcommand{\Bavcig}{\avbc_0}																	
\newcommand{\BopC}{\mathbb{C}}																
\newcommand{\BinC}{\mathsf{R}_{out}}									
\newcommand{\BinR}{\mathsf{R}_{in}}									

\newcommand{\BrCcompound}{\BopC^{\rstarC}\hspace{-0.1cm}(\Bcompound)}
\newcommand{\BrCcompoundP}{\BopC^{\rstarC}\hspace{-0.1cm}(\BcompoundP)}

\newcommand{\BrCav}{\BopC^{\rstarC}\hspace{-0.1cm}(\avbc)}
\newcommand{\BrCavig}{\BopC^{\rstarC}\hspace{-0.1cm}(\Bavcig)}
\newcommand{\BrICav}{\inR_{out}^{\rstarC}(\avbc)}

\newcommand{\BrIRavig}{\mathsf{R}^{\rstarC}_{0,in}}

\newcommand{\BCrp}{\BopC(\Brp)}
\newcommand{\BCrpnc}{\BopC(\Brpnc)}
\newcommand{\BCcompound}{\BopC(\Bcompound)}
\newcommand{\BCavc}{\BopC(\avbc)}
\newcommand{\BCavcig}{\BopC(\Bavcig)}
\newcommand{\BICrp}{\mathsf{C}(\Brp)}
\newcommand{\BICrpnc}{\mathsf{C}(\Brpnc)}
\newcommand{\BICcompound}{\BinC(\Bcompound)}
\newcommand{\BIRcompound}{\BinR(\Bcompound)}

\newcommand{\BIRavc}{\inR_{in}^{\rstarC}(\avbc)}

\newcommand{\savbc}{\avbc_{D}}
\newcommand{\sBrp}{\tBset^q_D} 
\newcommand{\sBrpnc}{\tBset^q_{D,n-c}} 
\newcommand{\sBCavc}{\BopC(\savbc)}
\newcommand{\sBCrp}{\BopC(\sBrp)}
\newcommand{\sBCrpnc}{\BopC(\sBrpnc)}

\newcommand{\sBrCav}{\BopC^{\rstarC}\hspace{-0.1cm}(\savbc)}

\newcommand{\sBrICav}{\inR_{out}^{\rstarC}(\savbc)}
\newcommand{\sBIRavc}{\inR_{in}^{\rstarC}(\savbc)}

\newcommand{\sBICrp}{\inC(\sBrp)}
\newcommand{\sBICrpnc}{\inC(\sBrpnc)}
\newcommand{\sBCavcig}{\BopC(\avbc_{D,0})}

\newcommand{\dWbc}{W_{Y_2|X,S_2}}

\begin{document}

\maketitle

\begin{abstract} 
In this work, we study the arbitrarily varying broadcast channel (AVBC), when state information is available at the transmitter in a causal manner.  
We 
 establish inner and outer bounds on both the random code capacity region and the deterministic code capacity region with degraded message sets. 
 The capacity region is then determined for a class of channels satisfying a condition on the mutual informations between the strategy variables and the channel outputs.
 As an example, we consider the arbitrarily varying binary symmetric broadcast channel with correlated noises. We show cases where the condition holds, hence the capacity region is determined, and other cases where there is a gap between the bounds.
%
\end{abstract}


\blfootnote{
This research was supported by the Israel Science Foundation (grant No. 1285/16).
}

The arbitrarily varying channel (AVC) was first introduced by Blackwell \etal 
\cite{BBT:60p} to describe a communication channel with unknown statistics, that may change over time.
It  is often described as  communication in the presence of an 
adversary, or a \emph{jammer}, attempting to disrupt communication. 

The arbitrarily  varying broadcast channel (AVBC) without side information (SI) was first considered by Jahn \cite{Jahn:81p}, who derived an inner bound on the random code capacity region, namely the capacity region achieved by encoder and decoders  with a random experiment,  shared between the three parties. As indicated by Jahn, the arbitrarily varying broadcast channel inherits 
 some of the properties of its single user counterpart. In particular, the random code capacity region is not necessarily achievable using deterministic codes \cite{BBT:60p}. Furthermore, 
 Jahn showed that 
the deterministic code capacity region  either coincides with the 
random code capacity region or else, it has an empty interior \cite{Jahn:81p}. This phenomenon is an analogue of Ahlswede's dichotomy property \cite{Ahlswede:78p}. 
Then, in order to apply Jahn's inner bound, one has to verify whether the capacity region has non-empty interior or not.
As observed in \cite{HofBross:06p}, this can be resolved using the results of
Ericson  \cite{Ericson:85p} and Csisz{\'{a}}r and Narayan \cite{CsiszarNarayan:88p}.
Specifically, a necessary and sufficient condition for the capacity region to have a non-empty interior is that both user marginal channels are non-symmetrizable. 

Various models of interest
 involve SI available at the encoder.
In \cite{WinshtokSteinberg:06c}, 
 the  arbitrarily varying degraded broadcast channel with non-causal SI is addressed, 
using Ahlswede's Robustification and Elimination Techniques \cite{Ahlswede:86p}. 
The single user AVC with causal SI is addressed in the book  by Csisz{\'a}r and K{\"o}rner \cite{CsiszarKorner:82b}, while their approach is independent of Ahlswede's work.
 A straightforward application of
 Ahlswede's Robustification Technique (RT) 
 would violate 
 the causality requirement. 

 In this work, we study the AVBC with causal SI available at the encoder. 
We 
 extend  Ahlswede's Robustification and Elimination Techniques \cite{Ahlswede:78p,
Ahlswede:86p}, 
 originally  used in the setting of \emph{non}-causal SI. 
In particular, we derive a modified version of Ahlswede's RT, suited to the setting of causal SI. 
In a recent paper 
 by the authors \cite{PeregSteinberg:17c2},  a similar proof technique is applied to the arbitrarily varying \emph{degraded} broadcast channel with causal SI. Here, we generalize those results, and consider a \emph{general} broadcast channel with degraded message sets with causal SI.

We establish inner and outer bounds on the random code and deterministic code capacity regions.
Furthermore, we give conditions on the AVBC under which the bounds coincide, and the capacity region is determined. 
  As an example, we consider the arbitrarily varying binary symmetric broadcast channel with correlated noises. 
	We  show that in some cases,  the conditions hold and the capacity region is determined. Whereas, in other cases, there is a gap between the bounds.
%
%

\section{Definitions and Previous Results}
\label{sec:Bnotation}
\subsection{Notation}
We use the following notation conventions throughout. 
Calligraphic letters $\Xset,\Sset,\Yset,...$ are used for finite sets.
Lowercase letters $x,s,y,\ldots$  stand for constants and values of random variables, and uppercase letters $X,S,Y,\ldots$ stand for random variables.  
 The distribution of a random variable $X$ is specified by a probability mass function (pmf) 
	$P_X(x)=p(x)$ over a finite set $\Xset$. The set of all pmfs over $\Xset$ is denoted by $\pSpace(\Xset)$. 
 We use $x^j=(x_1,x_{2},\ldots,x_j)$ to denote  a sequence of letters from $\Xset$. 
 A random sequence $X^n$ and its distribution $P_{X^n}(x^n)=p(x^n)$ are defined accordingly. 
For a pair of integers $i$ and $j$, $1\leq i\leq j$, we define the discrete interval $[i:j]=\{i,i+1,\ldots,j \}$.

	\subsection{Channel Description}
	\label{subsec:BCs}
A state-dependent discrete memoryless broadcast channel 
$(\Xset\times\Sset,\bc,\Yset_1,\Yset_2)$ consists of a finite input alphabet $\Xset$, two finite output alphabets $\Yset_1$ and $\Yset_2$, a finite state 
alphabet $\Sset$, and a collection of conditional pmfs $\bc$. 
 The channel is memoryless without feedback, and therefore  
$
W_{Y_1^n,Y_2^n|X^n,S^n}(y_1^n,y_2^n|x^n,s^n)= \prod_{i=1}^n \bc(y_{1,i},y_{2,i}|x_i,s_i) 
$. 
The marginals $\sbc$ and $\wbc$ correspond to user 1 and user 2, respectively. 
Throughout, unless mentioned otherwise, it is assumed that the users have \emph{degraded message sets}. That is, the encoder sends a private message which is intended for user 1, and a public message which is intended for both users.
  For state-dependent broadcast channels with causal SI, the channel input at time $i\in[1:n]$ may depend on the sequence of past and present states $s^i$. 


The \emph{arbitrarily varying broadcast channel} (AVBC) is a discrete memoryless broadcast channel $\bc$  with a state sequence of unknown distribution,  not necessarily independent nor stationary. That is, $S^n\sim \qn(s^n)$ with an unknown joint pmf $\qn(s^n)$ over $\Sset^n$.
In particular, $\qn(s^n)$ can give mass $1$ to some state sequence $s^n$. 
We denote the AVBC 
 with causal SI by $\avbc=\{\bc\}$.

To analyze the AVBC with degraded message sets with causal SI, we consider the
 \emph{compound  broadcast channel}. 
Different models of  compound broadcast channels have been considered in the literature, as \eg in 
\cite{WLSSV:09p} and \cite{BPS:14a}.
Here, we define 
the compound  broadcast channel as a discrete memoryless broadcast channel with a discrete memoryless state, where the state distribution $q(s)$ is not known in exact, but rather belongs to a family of distributions $\Qset$, with $\Qset\subseteq \pSpace(\Sset)$. That is,  $S^n\sim\prod_{i=1}^n q(s_i)$, with an unknown pmf $q\in\Qset$ over $\Sset$.
 We denote the compound broadcast channel 
 with causal SI by $\Bcompound$. 

The random parameter broadcast channel  is a special case of a compound broadcast channel where the set $\Qset$ consists of a single distribution, \ie when the state sequence is memoryless and distributed according to a given state distribution $q(s)$. Hence, we denote the random parameter broadcast channel 
 with causal SI by $\Brp$. 

\begin{figure}[b] 

%
	%
\begin{center}
\hspace{-2cm}
\begin{tabular}{l|ccc}
\backslashbox{SI}{Channel}
 								  &$\;$ 	Random Parameter &$\;$ 	Compound $\;$ &  AVBC 	 
								 \\	[0.2cm]   \hline \\			[-0.2cm] 								
without SI				& 	--  		  &  $\Bcompoundig$					&	  $\Bavcig$					\\[0.2cm] 
causal SI					&	 $\Brp$   	&	 $\Bcompound$						&  $\avbc$							
\end{tabular}
\end{center}
\caption{ Notation of broadcast channel families. The columns correspond to the channel family, and the rows indicate the role of SI at the encoder. 
  }
\label{table:BchannelNotation}
\end{figure} 

In Figure~\ref{table:BchannelNotation}, we set the basic notation for the broadcast channel families that we consider.  
 The columns correspond to the channel families presented above, namely the random parameter broadcast channel, the compound broadcast channel and the AVBC. 
 The rows indicate the role of SI, namely the case of no SI and causal SI. 
In the first row, and throughout, we use the subscript `$0$' to indicate the case where SI is not available.


\subsection{Coding with Degraded Message Sets}
\label{subsec:Bcoding}
We introduce some preliminary definitions, starting with the definitions of a deterministic code and a random code for the AVBC $\avbc$ with degraded message sets with 
 causal SI. Note that in general,
 the term `code', unless mentioned otherwise, refers to a deterministic code.  

\begin{definition}[A code, an achievable rate pair and capacity region]
\label{def:Bcapacity}
A $(2^{nR_0},2^{nR_1},n)$ code for the AVBC $\avbc$ with degraded message sets with causal SI consists of the following;   
two  message sets $[1:2^{nR_0}]$ and $[1:2^{nR_1}]$,  where it is assumed throughout that $2^{nR_0}$ and $2^{nR_1}$ are integers, a sequence of $n$ encoding functions $\enci:  [1:2^{nR_0}]\times[1:2^{nR_1}]\times \Sset^i \rightarrow \Xset$, $ i\in [1:n]$,  and two  decoding functions, $\dec_1: \Yset_1^n\rightarrow [1:2^{nR_0}]\times[1:2^{nR_1}] $ and $\dec_2: \Yset_2^n\rightarrow [1:2^{nR_0}]$.    

At time $i\in [1:n]$, given a pair of messages $(m_0,m_1)\in$$ [1:2^{nR_0}]\times[1:2^{nR_1}]$ and a sequence $s^i$,
 the encoder transmits $x_i=\enci(m_0,m_1,s^i)$. The codeword is then given by 
\begin{align}
x^n= \encn(m_0,m_1,s^n) \triangleq \left(
\enc_1(m_0,m_1,s_1),\enc_2(m_0,m_1,s^2),\ldots,\enc_n(m_0,m_1,s^n)   \right) \,.
\end{align}
Decoder $1$ receives the channel output $y_1^n$, and finds an estimate for the message pair $(\hm_0,\hm_1)=g_1(y_1^n)$. Decoder 2 only   estimates  the common message with $\tm_0=g_2(y_2^n)$. 
We denote the code by $\code=\left(\encn(\cdot,\cdot,\cdot),\dec_1(\cdot),\dec_2(\cdot) \right)$.

 Define the conditional probability of error of $\code$ given a state sequence $s^n\in\Sset^n$ by  
\begin{align}
\label{eq:Bcerr}
\cerr(\code)
=&\frac{1}{2^{ n(R_0+R_1 )}}\sum_{m_0=1}^{2^ {nR_0}} \sum_{m_1=1}^{2^ {nR_1}}
\sum_{
\Dset(m_0,m_1)^c} 
 \nBC(y_1^n,y_2^n|\encn(m_0,m_1,s^n),s^n) \,,
\end{align}
where 
\begin{align}
\Dset(m_0,m_1)\triangleq
\big\{\, (y_1^n,y_2^n)\in\Yset_1^n\times\Yset_2^n:  \dec_1(y_1^n)= (m_0,m_1) \,,\; \dec_2(y_2^n)= m_0 \,\big\} \,.
\end{align}
Now, define the average probability of error of $\code$ for some distribution $\qn(s^n)\in\pSpace(\Sset^n)$, 
\begin{align}
\err(\qn,\code)
=\sum_{s^n\in\Sset^n} \qn(s^n)\cdot\cerr(\code) \,.
\end{align}
We say that $\code$ is a 
$(2^{nR_0},2^{nR_1},n,\eps)$ code for the AVBC $\avbc$ if it further satisfies 
\begin{align}
\label{eq:Berr}
 \err(\qn,\code)\leq \eps \,,\quad\text{for all $\qn(s^n)\in\pSpace(\Sset^n)\,$.} 
\end{align} 

  We say that a rate pair $(R_0,R_1)$ is achievable
if for every $\eps>0$ and sufficiently large $n$, there exists a  $(2^{nR_0},2^{nR_1},n,\eps)$ code. The operational capacity region is defined as the closure of the set of achievable rate pairs and it is denoted by $\BCavc$. 
 We use the term `capacity region' referring to this operational
meaning, and in some places we call it the deterministic code capacity region in order to emphasize that achievability is measured with respect to  deterministic codes.  
\end{definition} 

We proceed now to define the parallel quantities when using stochastic-encoder stochastic-decoders triplets with common randomness.
The codes formed by these triplets are referred to as random codes. 

\begin{definition}[Random code]
\label{def:BcorrC} 
A $(2^{nR_0},2^{nR_1},n)$ random code for the AVBC $\avbc$ consists of a collection of 
$(2^{nR_0},2^{nR_1},n)$ codes $\{\code_{\gamma}=(\encn_\gamma,\dec_{1,\gamma},\dec_{2,\gamma})\}_{\gamma\in\Gamma}$, along with a probability distribution $\mu(\gamma)$ over the code collection $\Gamma$. 
We denote such a code by $\gcode=(\mu,\Gamma,\{\code_{\gamma}\}_{\gamma\in\Gamma})$.

Analogously to the deterministic case,  a $(2^{nR_0},2^{nR_1},n,\eps)$ random code has the additional requirement
\begin{align}
 \err(\qn,\gcode)=\sum_{\gamma\in\Gamma} \mu(\gamma)\err(q,
\code_\gamma)\leq \eps \,,\;\text{for all $\qn(s^n)\in\pSpace(\Sset^n)$} & \,. \qquad
\end{align}
The capacity region achieved by random codes is denoted by $\BrCav$, and it 
 is referred to as the random code capacity region.
\end{definition}
 
%

Next, we write the definition of superposition coding \cite{Bergmans:73p} using  Shannon strategies \cite{Shannon:58p}. See also \cite{Steinberg:05p}, and the discussion after Theorem 4 therein. Here, we refer to such codes as Shannon strategy codes.
\begin{definition}[Shannon strategy codes] 
\label{def:BStratCode}
A $(2^{nR_0},2^{nR_1},n)$ Shannon strategy code for the AVBC $\avbc$ with degraded message sets with causal SI is a $(2^{nR_0},2^{nR_1},n)$ 
 code 
with an encoder that is composed of two strategy sequences
\begin{align}
u_0^n :& [1:2^{nR_0}] \rightarrow \Uset_0^n \,, \\
u_1^n :& [1:2^{nR_0}]\times [1:2^{nR_1}] \rightarrow \Uset_1^n \,, 
\end{align}
 and an encoding function $\encs(u_0,u_1,s)$, where $\encs:\Uset_0\times\Uset_1\times\Sset\rightarrow \Xset$,  
as well as a pair of decoding functions
$
\dec_1: \Yset_1^n\rightarrow [1:2^{nR_0}]\times [1:2^{nR_1}] 
$ and 
$
\dec_2: \Yset_2^n\rightarrow [1:2^{nR_0}] 
$. The codeword is then given by 
\begin{align}
\label{eq:BStratEnc}
x^n=\encs^n(u_0^n(m_0),u_1^n(m_0,m_1),s^n)\triangleq \big[\, \encs(u_{0,i}^n(m_0),u_{1,i}^n(m_0,m_1),s_i) \,\big]_{i=1}^n \,.
\end{align}
We denote the code by $\code=\left(u_0^n,u_1^n,\encs,\dec_1,\dec_2 \right)$.
\end{definition}


\subsection{In the Absence of Side Information -- Inner Bound } 
\label{sec:BNOsi}
In this subsection, we briefly review known results for the case where the state is not known to the encoder or the decoder, \ie SI is not available. 

Consider a given AVBC with degraded message sets without SI, which we denote by $\Bavcig$.  
Let
\begin{align}
\label{eq:BrCavc0def}
\BrIRavig\triangleq
\bigcup_{p(x,u)} \bigcap_{q(s)}    
\left\{
\begin{array}{lrl}
(R_0,R_1) \,:\; & R_0 &\leq   I_q(U;Y_2) \,, \\
								& R_1 &\leq   I_q(X;Y_1|U) \,,\\
								& R_0+R_1 &\leq   I_q(X;Y_1)  \\	
\end{array}
\right\} 
\end{align}  
In \cite[Theorem~2]{Jahn:81p}, Jahn introduced an inner bound for 
 the arbitrarily varying \emph{general} broadcast channel. In our case, 
with 
 degraded message sets, 
 Jahn's inner bound reduces to the following. 

\begin{theorem}[Jahn's Inner Bound \cite{Jahn:81p}] 
\label{Btheo:avcC0R}
Let $\Bavcig$ be an AVBC with degraded message sets  without SI. Then, $\BrIRavig$ is an achievable rate region using random codes over $\Bavcig$, \ie
\begin{align}
\BrCavig \supseteq \BrIRavig \,.
\end{align}
\end{theorem}


Now we move to the deterministic code capacity region. 
\begin{theorem}[Ahlswede's Dichotomy \cite{Jahn:81p}] 
\label{theo:BavcC0}
The capacity region of   an AVBC $\Bavcig$ with degraded message sets  without SI either coincides with the random code capacity region or else, its interior is empty.
That is, 
$\BCavcig = \BrCavig$ 
 or else, 
$\interior{\BCavcig}=\emptyset$. 
\end{theorem} 
By Theorem~\ref{Btheo:avcC0R} and Theorem~\ref{theo:BavcC0}, we have that $\BrIRavig$ is an achievable rate region, 
if the interior of the capacity region is non-empty. That is,
$
\BCavcig \supseteq \BrIRavig 
$, if 
$ \interior{\BCavcig}\neq\emptyset$. 
\begin{theorem} [see \cite{
Ericson:85p,CsiszarNarayan:88p,HofBross:06p}]\label{theo:Bsymm0}
For an AVBC $\Bavcig$ without SI, the interior of the capacity region is non-empty, \ie  $\interior{\BCavcig}\neq\emptyset$, if and only if the marginals $\sbc$ and $\wbc$ are \emph{not} symmetrizable. 
\end{theorem}


\section{Main Results}
\label{sec:Bres}
We present our results on the compound broadcast channel and the AVBC with degraded message sets with causal SI. 

\subsection{The Compound Broadcast Channel with Causal SI}  
\label{sec:Bcompound}
We now consider the case where the encoder has access to the state sequence in a causal manner, \ie
 the encoder has $S^i$. 
%

\subsubsection{Inner Bound}
 First, we provide an achievable rate region for the compound broadcast channel with degraded message sets with causal SI. Consider a given compound broadcast channel $\Bcompound$ with causal SI. 
Let 
\begin{align}  
\label{eq:BIRcompound} 
\BIRcompound \triangleq\bigcup_{p(u_0,u_1),\,\encs(u_0,u_1,s)}\, \bigcap_{q(s)\in\Qset} 
\left\{
\begin{array}{lrl}
(R_0,R_1) \,:\; & R_0 		\leq&   I_q(U_0;Y_2) \,, \\
								& R_1 		\leq&   I_q(U_1;Y_1|U_0) \,,\\
								& R_0+R_1	\leq&   I_q(U_0,U_1;Y_1)
\end{array}
\right\} 
\end{align}
subject to $X=\encs(U_0,U_1,S)$, where $U_0$ and $U_1$ are auxiliary random variables, independent of $S$, and the union is over the pmf $p(u_0,u_1)$ and the set of all functions $\encs:\Uset_0\times\Uset_1\times\Sset\rightarrow\Xset$.
This can also be expressed as
\begin{align}
\label{eq:BIRcompoundEQ}
\BIRcompound=\bigcup_{p(u_0,u_1),\,\encs(u_0,u_1,s)} 
\left\{
\begin{array}{lrl}
(R_0,R_1) \,:\; & R_0 	\leq&   \inf_{q\in\Qset} I_q(U_0;Y_2) \,, \\
								& R_1 	\leq&   \inf_{q\in\Qset} I_q(U_1;Y_1|U_0)\,,\\
								&R_0+R_1\leq& 	\inf_{q\in\Qset} I_q(U_0,U_1;Y_1)	
\end{array}
\right\}\,.
\end{align} 


\begin{lemma}
\label{lemm:BcompoundLowerB}
Let $\Bcompound$ be a compound broadcast channel with degraded message sets with causal SI available at the encoder. Then, $\BIRcompound$ is an achievable rate region for $\Bcompound$, \ie 
\begin{align}
\BCcompound \supseteq \BIRcompound \,.
\end{align}
Specifically, if $(R_0,R_1)\in\BIRcompound$, then for some $a>0$ and sufficiently large $n$, there exists a $(2^{nR_0},2^{nR_1},n,e^{-an})$ Shannon strategy code over the compound broadcast channel $\Bcompound$ with degraded message sets with causal SI.
\end{lemma}
The proof of Lemma~\ref{lemm:BcompoundLowerB} is given in Appendix~\ref{app:BcompoundLowerB}.

%
%

\subsubsection{The Capacity Region}
\label{subsec:BcvC}
We determine the capacity region of the compound broadcast channel $\Bcompound$ with degraded message sets with causal SI available at the encoder. 
In addition, we give a condition, for which the inner bound in Lemma~\ref{lemm:BcompoundLowerB} coincides with the capacity region.
Let
\begin{align}  
\label{eq:BcvCI}
\BICcompound \triangleq  \bigcap_{q(s)\in\Qset} 
\bigcup_{p(u_0,u_1),\,\encs(u_0,u_1,s)}
\left\{
\begin{array}{lrl}
(R_0,R_1) \,:\; & R_0 	\leq&   I_q(U_0;Y_2) \,, \\
								& R_1 	\leq&   I_q(U_1;Y_1|U_0) \,\\
								&R_0+R_1\leq&   I_q(U_0,U_1;Y_1)	
\end{array}
\right\}
 \,.
\end{align}
 
Now, our condition is defined in terms of the following.
\begin{definition}
\label{def:Bcompoundachieve} 
We say that a function $\encs:\Uset_0\times\Uset_1\times\Sset\rightarrow\Xset$ and a set $\Dset\subseteq\pSpace(\Uset_0\times\Uset_1)$ achieve both $\BIRcompound$ and $\BICcompound$ if 
\begin{subequations}
\label{eq:Bcompoundachieve} 
\begin{align}  
\label{eq:BIRcompoundachieve} 
\BIRcompound =\bigcup_{p(u_0,u_1)\in\Dset}\, \bigcap_{q(s)\in\Qset} 
\left\{
\begin{array}{lrl}
(R_0,R_1) \,:\; & R_0	 	\leq&   I_q(U_0;Y_2) \,, \\
								& R_1	 	\leq&   I_q(U_1;Y_1|U_0) \,,\\
								&R_0+R_1\leq&   I_q(U_0,U_1;Y_1)	
\end{array}
\right\} \,,
\intertext{and}
\label{eq:BICcompoundachieve} 
\BICcompound = \bigcap_{q(s)\in\Qset}\, \bigcup_{p(u_0,u_1)\in\Dset} 
\left\{
\begin{array}{lrl}
(R_0,R_1) \,:\; & R_0 	\leq&   I_q(U_0;Y_2) \,, \\
								& R_1 	\leq&   I_q(U_1;Y_1|U_0) \,,\\
								&R_0+R_1\leq&   I_q(U_0,U_1;Y_1)	
\end{array}
\right\} \,,
\end{align}
\end{subequations}
subject to $X=\encs(U_0,U_1,S)$. That is, the unions in  (\ref{eq:BIRcompound}) and (\ref{eq:BcvCI}) 
  can be restricted to the particular
function $\encs(u_0,u_1,s)$ and set of strategy distributions $\Dset$.
\end{definition}
Observe that by Definition~\ref{def:Bcompoundachieve},  given a function $\encs(u_0,u_1,s)$, if
a set $\Dset$ achieves both $\BIRcompound$ and $\BICcompound$, then every set $\Dset'$ with $\Dset\subseteq\Dset'\subseteq\pSpace(\Uset_0\times\Uset_1)$ achieves those regions, 
 and in particular,  $\Dset'=\pSpace(\Uset_0\times\Uset_1)$. Nevertheless, the condition defined below requires a certain property  that may hold for $\Dset$, but not for $\Dset'$. 

\begin{definition} 
\label{def:sCondQ}
Given a convex set $\Qset$ of state distributions, define Condition $\sCondQ$ by the following;
for some $\encs(u_0,u_1,s)$ and $\Dset$ that achieve both $\BIRcompound$ and $\BICcompound$,
 there exists $q^*\in\Qset$ which minimizes the mutual informations $I_q(U_0;Y_2)$, $I_q(U_1;Y_1|U_0)$, and $I_q(U_0,U_1;Y_1)$, for all $p(u_0,u_1)\in\Dset$, 
\ie
\begin{IEEEeqnarray}{ll}
\sCondQ \,:\; &\text{For some $q^*\in\Qset$,}  \nonumber\\
 &q^*=\arg\min_{q\in\Qset} I_q(U_0;Y_2)=\arg\min_{q\in\Qset} I_q(U_1;Y_1|U_0)
=\arg\min_{q\in\Qset} I_q(U_0,U_1;Y_1) \,, \quad
\nonumber\\
& \forall p(u_0,u_1)\in\Dset 
 \,.
\label{eq:TQ}
\end{IEEEeqnarray}
\end{definition}
Intuitively, when Condition $\sCondQ$ holds, there exists a single jamming strategy $q^*(s)$ which is worst for both users simultaneously. That is, there is no tradeoff for the jammer. As the optimal jamming strategy is unique, this eliminates ambiguity for the users as well.

\begin{theorem}
\label{theo:BcvC} 
Let $\Bcompound$ be a compound broadcast channel  with causal SI available at the encoder. Then,
\begin{enumerate}[{1)}]
\item
the capacity region of $\Bcompound$ follows 
\begin{align} 
\label{eq:BcvC}
\BCcompound= \BICcompound  \,,\;\,\text{if $\interior{\BCcompound}\neq\emptyset$}\,, 
\end{align}
and it is identical to the corresponding random code capacity region, \ie  $\BrCcompound=\BCcompound$ if $\interior{\BCcompound}\neq\emptyset$.
\item
Suppose that $\Qset\subseteq\pSpace(\Sset)$ is a convex set of state distributions. If Condition $\sCondQ$ holds, the capacity region of  $\Bcompound$ is given by 
\begin{align}
\BCcompound=\BIRcompound=\BICcompound \,,
\end{align}
and it is identical to the corresponding random code capacity region, \ie $\BrCcompound=\BCcompound$. 
\end{enumerate}
%
\end{theorem}
The proof of Theorem~\ref{theo:BcvC} is given in Appendix~\ref{app:BcvC}. 
Regarding part 1,  we note that when $\interior{\BCcompound}=\emptyset$, then the inner bound 
$\BIRcompound$ has an empty interior as well (see (\ref{eq:BIRcompoundEQ})). Thus,
 $\interior{\BIRcompound}\neq\emptyset$ is also a sufficient condition for $\BCcompound=\BICcompound$. 
%
\subsubsection{The Random Parameter Broadcast Channel with Causal SI}
Consider the random parameter broadcast channel with causal SI. Recall that this is simply a special case of a compound broadcast channel, where the set of state distributions consists of a single member, \ie $\Qset=\{q(s)\}$. Then, let
\begin{align}
\label{eq:BICrp} 
\BICrp \triangleq&  \bigcup_{p(u_0,u_1),\encs(u_0,u_1,s)} 
\left\{
\begin{array}{lrl}
(R_0,R_1) \,:\; & R_0 	\leq&   I_q(U_0;Y_2) \,, \\
								& R_1 	\leq&   I_q(U_1;Y_1|U_0) \,,\\
								&R_0+R_1\leq&   I_q(U_0,U_1;Y_1)	
\end{array}
\right\} \,,
\end{align}
with
\begin{align}
 |\Uset_0|\leq& |\Xset| |\Sset|+2 \,,\; 
|\Uset_1|\leq |\Xset| |\Sset|( |\Xset| |\Sset|+2) \,.
\label{eq:BICrpAlph}
\end{align}
\begin{theorem}
\label{theo:BCrp}
The capacity region of the random parameter broacast channel $\Brp$ with degraded message sets with causal SI is given by 
\begin{align}
\BCrp=\BICrp \,.
\end{align}
\end{theorem}
Theorem~\ref{theo:BCrp} is proved in Appendix~\ref{app:BCrp}. 
\subsection{The AVBC with Causal SI}
We give inner and outer bounds, on the random code capacity region and the deterministic code capacity region, for the 
AVBC $\avbc$ with degraded message sets with causal SI. We also provide conditions, for which the inner bound coincides with the outer bound.
\subsubsection{Random Code Inner and Outer Bounds}
Define 
\begin{align}
\label{eq:BIRcompoundP}  
&\BIRavc \triangleq \BIRcompound\bigg|_{\Qset=\pSpace(\Sset)}
\,,\;
\BrICav\triangleq \BICcompound\bigg|_{\Qset=\pSpace(\Sset)}
\,,
\intertext{and}
&\sCond=\sCond^{\Qset}\Big|_{\Qset=\pSpace(\Sset)} \,.
\label{eq:sCondShort}
\end{align}

\begin{theorem}
\label{theo:Bmain}
Let $\avbc$ be an AVBC with degraded message sets with causal SI available at the encoder. Then,  
\begin{enumerate}[{1)}]
\item
the random code capacity region of $\avbc$ is bounded by
\begin{align}
\BIRavc \subseteq \BrCav \subseteq \BrICav \,.
\end{align}
\item 
If Condition $\sCond$ holds, the random code capacity region of  $\avbc$ is given by
	\begin{align}
	\label{eq:BrCavTight}
	\BrCav=\BIRavc=\BrICav \,.
	\end{align}
\end{enumerate}
\end{theorem}
The proof of Theorem~\ref{theo:Bmain} is given in Appendix~\ref{app:Bmain}. 

Before we proceed to the deterministic code capacity region, we need one further result.
The following lemma is a restatement of a result from \cite{Ahlswede:78p}, stating that a polynomial size of the code collection $\{\code_\gamma\}$ is sufficient. This result is a key observation in Ahlswede's Elimination Technique (ET), presented in \cite{Ahlswede:78p}, and it is significant for the deterministic code analysis.   
\begin{lemma} 
\label{lemm:BcorrSizeC}  
Consider a given  
 $(2^{nR_0},2^{nR_1},n,\eps_n)$ random 
code $\code^\Gamma=(\mu,\Gamma,\{\code_\gamma\}_{\gamma\in\Gamma})$
 for the AVBC $\avbc$, 
where $\lim_{n\rightarrow\infty} \eps_n=0$. 
Then, for every $0<\alpha<1$ and sufficiently large $n$, there exists a $(2^{nR_0},2^{nR_1},n,\alpha)$ random 
 code $(\mu^*,\Gamma^*,\{\code_{\gamma}
\}_{\gamma\in\Gamma^*})$ with the following properties:
\begin{enumerate}
 \item 
The size of the code collection is bounded by
 $
 |\Gamma^*|\leq n^2 
$. 
\item
\label{item:Bsubset}
 The code collection is a subset of the original code collection, \ie 
$
\Gamma^*\subseteq \Gamma 
$. 
\item
 The distribution $\mu^*$ 
 is uniform, \ie 
$
\mu^*(\gamma)=\frac{1}{|\Gamma^*|} 
$, 
for $\gamma\in\Gamma^*$. 
\end{enumerate} 
\end{lemma} 
The proof of Lemma~\ref{lemm:BcorrSizeC} follows 
 the same lines  as in \cite[Section 4]{Ahlswede:78p} (see also \cite{Jahn:81p,WinshtokSteinberg:06c}).
 For completeness, we give the proof in Appendix~\ref{app:BET}.

\subsubsection{Deterministic Code Inner and Outer Bounds}
The next theorem characterizes the deterministic code capacity region, which demonstrates a dichotomy property. 
\begin{theorem}
\label{theo:BcorrTOdetC}
The capacity region of an AVBC $\avbc$ with degraded message sets with causal SI either coincides with the random code capacity region or else, it has an empty interior. 
That is, 
$\BCavc = \BrCav$ 
 or else, 
$ \interior{\BCavc}=\emptyset$. 
\end{theorem}
The proof of Theorem~\ref{theo:BcorrTOdetC} is given in Appendix~\ref{app:BcorrTOdetC}.
Let $U=(U_0,U_1)$, hence $\Uset=\Uset_0\times\Uset_1$.
For every pair of functions $\encs:\Uset\times\Sset\rightarrow\Xset$ and $\encs':\Uset_0\times\Sset\rightarrow\Xset$, define the DMCs  $V_{Y_1|U,S}^{\encs}$ and $V_{Y_2|U_0,S}^{\encs'}$ specified by 
\begin{subequations}
\label{eq:BchannelUY1Y2}
\begin{align}
&V_{Y_1|U,S}^{\encs}(y_1|u,s)=W_{Y_1|X,S}(y_1|\encs(u,s),s) \,, \\
&V_{Y_2|U_0,S}^{\encs'}(y_2|u_0,s)=W_{Y_2|X,S}(y_2|\encs'(u_0,s),s) \,,
\end{align}
\end{subequations}
 respectively.
%
%
\begin{coro}
\label{coro:BmainDbound}
The capacity region of $\avbc$ is bounded by
\begin{align}
&\BCavc \supseteq \BIRavc \,,\;\text{if}\;\, \interior{\BCavc}\neq \emptyset \,, \label{eq:BmainInner} \\
&\BCavc \subseteq \BrICav \,. \label{eq:BmainOuter}
\end{align}
Furthermore,
if  $V_{Y_1|U,S}^{\encs}$ and $V_{Y_2|U_0,S}^{\encs'}$ are non-symmetrizable for some  $\encs:\Uset\times\Sset\rightarrow\Xset$ and $\encs':\Uset_0\times\Sset\rightarrow\Xset$, and Condition $\sCond$ holds, then 
 $\,\BCavc=\BIRavc=\BrICav$.
\end{coro}
The proof of Corollary~\ref{coro:BmainDbound} is given in Appendix~\ref{app:BmainDbound}.

\section{Degraded Broadcast Channel with Causal SI}
In this section, we consider the special case of an arbitrarily varying degraded broadcast channel (AVDBC) with causal SI, when user 1 and user 2 have private messages.

\subsection{Definitions}
\label{subsec:Bdegdef}
We consider a \emph{degraded broadcast channel} (DBC), which is a special case of the general broadcast channel described in the previous sections.
%
 Following the definitions by \cite{Steinberg:05p}, a state-dependent broadcast channel $\bc$ is said to be physically degraded if it can be expressed as 
\begin{align}
\label{eq:Bdegraded}
\bc(y_1,y_2|x,s)=W_{Y_1|X,S}(y_1|x,s)\cdot 
p(y_2|y_1) \;,
\end{align}
 \ie $(X,S)\Cbar Y_1 \Cbar Y_2$ form a Markov chain. 
User 1 is then referred to as the \emph{stronger} user, whereas user 2 is referred to as the \emph{weaker} user. 
More generally, a broadcast channel is said to be stochastically degraded if  
 $\wbc(y_2|x,s)=\sum_{y_1\in\Yset_1}W_{Y_1|X,S}(y_1|$ $x,s)\cdot$ $ 
\tp(y_2|y_1)$
for some conditional distribution $\tp(y_2|y_1)$. We note that the definition of degradedness here is stricter than the definition in \cite[Remark IIB5]
{Jahn:81p}. 
Our results apply to both the physically degraded and the stochastically degraded broadcast channels. 
Thus, for our purposes,
 there is no need to distinguish between the two, and we simply say that the broadcast channel is degraded.
We use the notation $\savbc$ for an AVDBC with causal SI.

We consider the case where the users have private messages.
A deterministic code and a random code for the AVDBC $\savbc$ with 
 causal SI are then defined as follows. 

\begin{definition}[A private-message code, an achievable rate pair and capacity region]
\label{def:sBcapacity}
A $(2^{nR_1},2^{nR_2},n)$ private-message code for the AVDBC $\savbc$ 
with causal SI consists of the following;   
two  message sets $[1:2^{nR_1}]$ and $[1:2^{nR_2}]$,  
where it is assumed throughout that $2^{nR_1}$ and $2^{nR_2}$ are integers,
a set of $n$ encoding functions 
$\enci:  [1:2^{nR_1}]\times[1:2^{nR_2}]\times \Sset^i \rightarrow \Xset$,   
$ i\in [1:n]$, 
 and two  decoding functions,
$
\dec_1: \Yset_1^n\rightarrow [1:2^{nR_1}]  
$ 
and 
$
\dec_2: \Yset_2^n\rightarrow [1:2^{nR_2}]   
$. 

At time $i\in [1:n]$, given a pair of messages $m_1\in [1:2^{nR_1}]$ and $m_2\in[1:2^{nR_2}]$ and a sequence $s^i$,
 the encoder transmits $x_i=\enci(m_1,m_2,s^i)$. The codeword is then given by 
\begin{align}
x^n= \encn(m_1,m_2,s^n) \triangleq \left(
\enc_1(m_1,m_2,s_1),\enc_2(m_1,m_2,s^2),\ldots,\enc_n(m_1,m_2,s^n)   \right) \,.
\end{align}
Decoder $k$ receives the channel output $y_k^n$, for $k=1,2.$,  and finds an estimate for the $k^{\text{th}}$ message,
$\hm_k=\dec_k(y_k^n)$. Denote the code by $\code=\left(\encn(\cdot,\cdot,\cdot),\dec_1(\cdot),\dec_2(\cdot) \right)$.

 Define the conditional probability of error of $\code$ given a state sequence $s^n\in\Sset^n$ by  
\begin{align}
\label{eq:Bcerr1}
\cerr(\code)
=&\frac{1}{2^{ n(R_1+R_2 )}}\sum_{m_1=1}^{2^ {nR_1}} \sum_{m_2=1}^{2^ {nR_2}}
\sum_{
\Dset(m_1,m_2)^c} 
 \nBC(y_1^n,y_2^n|\encn(m_1,m_2,s^n),s^n) \,,
\end{align}
where 
\begin{align}
\Dset(m_1,m_2)\triangleq
\big\{\, (y_1^n,y_2^n)\in\Yset_1^n\times\Yset_2^n:  \dec_1(y_1^n)= m_1 \,,\; \dec_2(y_2^n)= m_2 \,\big\} \,.
\end{align}
We say that $\code$ is a 
$(2^{nR_1},2^{nR_2},n,\eps)$ code for the AVDBC $\avbc$ if it further satisfies 
\begin{align}
\label{eq:Berr1}
 \err(\qn,\code)=\sum_{s^n\in\Sset^n} \qn(s^n)\cdot\cerr(\code)\leq \eps \,,\quad\text{for all $\qn(s^n)\in\pSpace(\Sset^n)\,$.} 
\end{align} 
An achievable private-message rate pair $(R_1,R_2)$ and the 
 capacity region $\sBCavc$ are defined as usual.
\end{definition} 

We proceed now to define the parallel quantities when using stochastic-encoder stochastic-decoders triplets with common randomness.

\begin{definition}[Random code]
\label{def:sBcorrC} 
A $(2^{nR_1},2^{nR_2},n)$ private-message random code for the AVDBC $\savbc$ consists of a collection of 
$(2^{nR_1},2^{nR_2},n)$ codes $\{\code_{\gamma}=(\encn_\gamma,\dec_{1,\gamma},\dec_{2,\gamma})\}_{\gamma\in\Gamma}$, along with a probability distribution $\mu(\gamma)$ over the code collection $\Gamma$. 

Analogously to the deterministic case,  a $(2^{nR_1},2^{nR_2},n,\eps)$ random code has the additional requirement
\begin{align}
 \err(\qn,\gcode)=\sum_{\gamma\in\Gamma} \mu(\gamma)\err(q,
\code_\gamma)\leq \eps \,,\;\text{for all $\qn(s^n)\in\pSpace(\Sset^n)$} & \,. 
\end{align}
The private-message capacity region achieved by random codes is denoted by $\sBrCav$, and it 
 is referred to as the random code capacity region.
\end{definition}
 
By standard arguments,  a private-message rate pair $(R_1,R_2)$ is achievable for the AVDBC $\savbc$  if and only if $(R_0,R_1)$ is achievable with degraded message sets, with $R_0=R_2$.
This immediately implies the following results.

\subsection{Results}
The results in this section are a straightforward consequence of the results in Section~\ref{sec:Bres}.
\subsubsection{Random Code Inner and Outer Bounds}
Define 
\begin{align}
\label{eq:BIRcompoundP1} 
&\sBIRavc \triangleq\bigcup_{p(u_1,u_2),\,\encs(u_1,u_2,s)}\; \bigcap_{q(s)} \,
\left\{
\begin{array}{lrl}
(R_1,R_2) \,:\; & R_2 &\leq   I_q(U_2;Y_2) \,, \\
								& R_1 &\leq   I_q(U_1;Y_1|U_2) 
\end{array}
\right\} \,,
\intertext{and}
\label{eq:BrICav1}
&\sBrICav\triangleq \bigcap_{q(s)}\;  \bigcup_{p(u_0,u_1),\,\encs(u_0,u_1,s)} \,
\left\{
\begin{array}{lll}
(R_1,R_2) \,:\; & R_2 &\leq   I_q(U_2;Y_2) \,, \\
								& R_1 &\leq   I_q(U_1;Y_1|U_2)
\end{array}
\right\} \,.
\end{align}

Now, we define a condition in terms of the following.
\begin{definition}
\label{def:Bachieve1} 
We say that a function $\encs:\Uset_1\times\Uset_2\times\Sset\rightarrow\Xset$ and a set $\Dset^{\rstarC}\subseteq\pSpace(\Uset_1\times\Uset_2)$ achieve both $\sBIRavc$ and $\sBrICav$ if 
\begin{subequations}
\label{eq:Bachieve1} 
\begin{align}  
\label{eq:BIRachieve1} 
\sBIRavc =\bigcup_{p(u_0,u_1)\in\Dset^{\;\,\rstarC}}\, \bigcap_{q(s)} 
\left\{
\begin{array}{lll}
(R_1,R_2) \,:\; & R_2 &\leq   I_q(U_2;Y_2) \,, \\
								& R_1 &\leq   I_q(U_1;Y_1|U_2)
\end{array}
\right\} \,,
\intertext{and}
\label{eq:BICachieve1} 
\sBrICav = \bigcap_{q(s)}\, \bigcup_{p(u_0,u_1)\in\Dset^{\;\,\rstarC}} 
\left\{
\begin{array}{lll}
(R_1,R_2) \,:\; & R_2 &\leq   I_q(U_2;Y_2) \,, \\
								& R_1 &\leq   I_q(U_1;Y_1|U_2)
\end{array}
\right\} \,,
\end{align}
\end{subequations}
subject to $X=\encs(U_1,U_2,S)$. That is, the unions in (\ref{eq:BIRcompoundP1}) and (\ref{eq:BrICav1}) can be restricted to the particular function $\encs(u_1,u_2,s)$ and set of strategy distributions $\Dset^{\rstarC}$.
\end{definition}

\begin{definition} 
\label{def:sCond1}
Define Condition $\sCond_D$ by the following;
for some $\encs(u_1,u_2,s)$ and $\Dset^{\rstarC}$ that achieve both $\sBIRavc$ and $\sBrICav$,
 there exists $q^*\in\pSpace(\Sset)$ which minimizes both $I_q(U_2;Y_2)$ and $I_q(U_1;Y_1|U_2)$, for all 
$p(u_1,u_2)\in\Dset^{\rstarC}$, 
\ie
\begin{IEEEeqnarray*}{ll}
\sCond_D \,:\; &\text{For some $q^*\in\pSpace(\Sset)$,}  \\ 
 &q^*=\arg\min_{q(s)} I_q(U_2;Y_2)=\arg\min_{q(s)} I_q(U_1;Y_1|U_2)
\quad
 \forall p(u_1,u_2)\in\Dset^{\rstarC} 
\,.
\end{IEEEeqnarray*}
\end{definition}

\begin{theorem}
\label{theo:Bmain1}
Let $\savbc$ be an AVDBC with causal SI available at the encoder. Then,  
\begin{enumerate}[{1)}]
\item
the random code capacity region of $\savbc$  is bounded by
\begin{align}
\sBIRavc \subseteq \sBrCav \subseteq \sBrICav \,.
\end{align}
\item 
If Condition $\sCond_D$ holds, the random code capacity region of  $\savbc$  is given by
	\begin{align}
	\label{eq:sBrCavTight}
	\sBrCav=\sBIRavc=\sBrICav \,.
	\end{align}
\end{enumerate}
\end{theorem}
Theorem~\ref{theo:Bmain1} is a straightforward consequence of Theorem~\ref{theo:Bmain}.

\subsubsection{Deterministic Code Inner and Outer Bounds}
The next theorem characterizes the deterministic code capacity region, which demonstrates a dichotomy property. 
\begin{theorem}
\label{theo:sBcorrTOdetC1}
The capacity region of an AVDBC $\savbc$ with causal SI either coincides with the random code capacity region or else, it has an empty interior. 
That is, 
$\sBCavc = \sBrCav$ 
 or else, 
$ \interior{\sBCavc}=\emptyset$. 
\end{theorem}
Theorem~\ref{theo:sBcorrTOdetC1} is a straightforward consequence of Theorem~\ref{theo:BcorrTOdetC}.
Now, Theorem~\ref{theo:Bmain1} and Theorem~\ref{theo:sBcorrTOdetC1} yield the following corollary. 
For every function  $\encs':\Uset_2\times\Sset\rightarrow\Xset$, define a DMC $V_{Y_2|U_2,S}^{\encs'}$ specified by 
\begin{align}
V_{Y_2|U_2,S}^{\encs'}(y_2|u_2,s)=&W_{Y_2|X,S}(y_2|\encs'(u_2,s),s) \,.
\end{align}
%
%
\begin{coro}
\label{coro:sBmainDbound}
The capacity region of $\savbc$ is bounded by
\begin{align}
&\sBCavc \supseteq \sBIRavc \,,\;\text{if}\;\, \interior{\sBCavc}\neq \emptyset \,, \label{eq:sBmainInner} \\
&\sBCavc \subseteq \sBrICav \,. \label{eq:sBmainOuter}
\end{align}
Furthermore,
if  $V_{Y_2|U_2,S}^{\encs'}$ is non-symmetrizable for some $\encs':\Uset_2\times\Sset\rightarrow\Xset$, and Condition $\sCond_D$ holds, then 
 $\,\sBCavc=\sBIRavc=\sBrICav$.
\end{coro}

%

\begin{center}
\begin{figure}[hbt]
        \centering
        \includegraphics[scale=0.52,trim={2.5cm 0 0 0},clip]{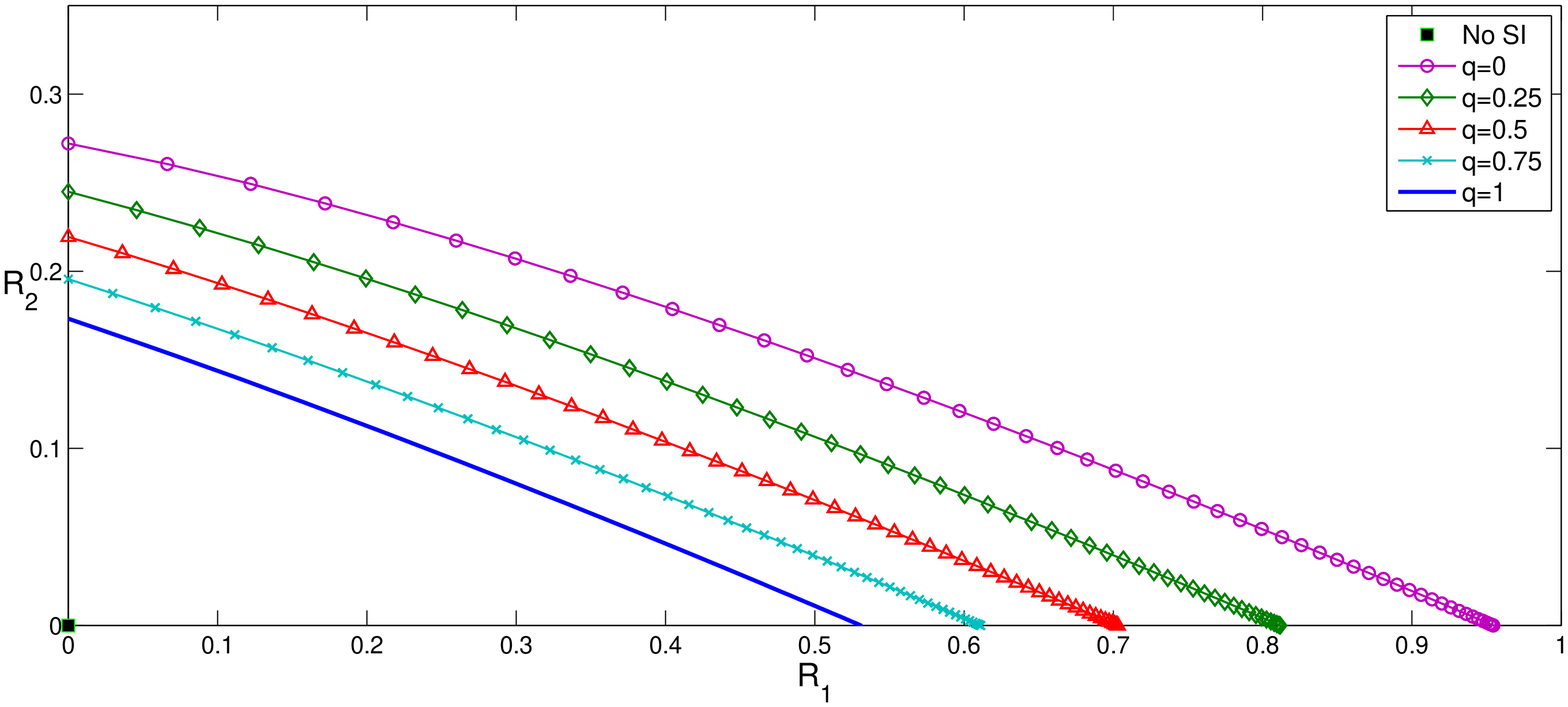}
        
\caption{The private-message capacity region of the AVDBC in Example~\ref{example:AVBSBC}, the arbitrarily varying binary symmetric broadcast channel. The area under the thick blue line is the  capacity region of the AVDBC $\savbc$ with causal SI, with $\theta_0=0.005$, 
$\theta_1=0.9$, and 
$\alpha=0.2$. The black square at the origin stands for the capacity region of the AVDBC $\avbc_{D,0}$ without SI, $\sBCavcig=\{(0,0)\}$.
The curves depict $\opC(\savbc^{q})$ for $q=0,0.25,0.5,0.75,1$, where the capacity region of $\savbc$ is given by $\sBCavc=\inR^{\,\;\rstarC}_{out}(\savbc)=\opC(\savbc^{q})$ for $q=1$ 
(see (\ref{eq:BrICav1})). 
  }
\label{fig:BBSC}
\end{figure}
\end{center}
\section{Examples}
\label{sec:exAVBSBC}
To illustrate the results above, we give the following examples.
In the first example, we consider an AVDBC and determine the private-message capacity region. 
Then, in the second example, we consider a non-degraded AVBC and determine the capacity region with degraded message sets.
\begin{example} 
\label{example:AVBSBC}
Consider an arbitrarily varying binary symmetric broadcast channel (BSBC), 
\begin{align}
Y_1=&X+Z_S \mod 2 \,, \nonumber\\
Y_2=&Y_1+K \mod 2 \,, \nonumber
\end{align}
where $X,Y_1,Y_2,S,Z_S,K$ are binary, with values in $\{0,1\}$. The additive noises are distributed according to 
\begin{align}
Z_s\sim& \text{Bernoulli}(\theta_s) \,,\; \text{for $s\in\{0,1\}$}\,, \nonumber\\
K\sim& \text{Bernoulli}(\alpha) \,, \nonumber
\end{align}
 with  $\theta_0\leq 1-\theta_1 \leq \frac{1}{2}$ and $\alpha<\frac{1}{2}$, where $K$ is independent of $(S,Z_S)$. 
It is readily seen the channel is  physically degraded.
Then, consider the case where user 1 and user 2 have private messages.

We have the following results.
 Define the binary entropy function $h(x)=-x\log x-(1-x)\log(1-x)$, for $x\in [0,1]$, with logarithm to base $2$.
The private-message capacity region of the arbitrarily varying BSBC $\avbc_{D,0}$ without SI is given by
\begin{align}
\label{eq:Bex1Cavcig}
\sBCavcig= \{(0,0)\} \,. 
\end{align}
The private-message capacity region of the arbitrarily varying BSBC $\savbc$ with causal SI is given by 
\begin{align}
\label{eq:Bex1Cavc}
\sBCavc= \bigcup_{0\leq \beta\leq 1}
\left\{
\begin{array}{lrl}
(R_1,R_2) \,:\; & R_2 &\leq   1-h(\alpha*\beta*\theta_1) \,, \\
								& R_1 &\leq   h(\beta*\theta_1)-h(\theta_1)
\end{array}
\right\}\,.
\end{align}
It will be seen in the achievability proof that the parameter $\beta$ is related to the distribution of  $U_1$, and thus the RHS of 
(\ref{eq:Bex1Cavc}) can be thought  of as a union over Shannon strategies.
The analysis is given in Appendix~\ref{app:AVBSBC}. 

It is shown in Appendix~\ref{app:AVBSBC} that Condition $\sCond_D$ holds and $\sBCavc=\sBIRavc=\sBrICav$.
Figure~\ref{fig:BBSC} provides a graphical interpretation. 
Consider a DBC $\bc$ with random parameters  with causal SI, governed by an i.i.d. state sequence, distributed according to $S\sim\text{Bernoulli}(q)$, for a given $0\leq q\leq 1$, and let $\opC(\savbc^{q})$ denote the corresponding capacity region.
 Then, the analysis shows that  Condition $\sCond_D$ implies that there exists $0\leq q^*\leq 1$ such that $\sBCavc=\opC(\savbc^{q^*})$, where $\opC(\savbc^{q^*})\subseteq \opC(\savbc^{q}) $ for every $0\leq q\leq 1$. Indeed,
looking at Figure~\ref{fig:BBSC}, it appears that the regions $\opC(\savbc^{q})$, for $0\leq q\leq 1$, form a well ordered set, hence 
 $\sBCavc=\opC(\savbc^{q^*})$ with $q^*=1$. 
%
\end{example}


Next, we consider an example of an AVBC which is not degraded in the sense defined above.
\begin{example} 
\label{example:AVBSBC2}
Consider a state-dependent binary symmetric broadcast channel (BSBC) with correlated noises, 
\begin{align}
Y_1=&X+Z_S \mod 2 \,, \nonumber\\
Y_2=&X+N_S \mod 2 \,, \nonumber
\end{align}
where $X,Y_1,Y_2,S,Z_S,N_S$ are binary, with values in $\{0,1\}$. The additive noises are distributed according to 
\begin{align}
Z_s\sim& \text{Bernoulli}(\theta_s) \,,\; 
N_s\sim \text{Bernoulli}(\eps_s) \,,\; \text{for $s\in\{0,1\}$}\,, \nonumber
\end{align}
 where $S,Z_0,Z_1,N_0,N_1$ are independent random variables, with $\theta_0\leq \eps_0\leq \frac{1}{2}$ and $ \frac{1}{2}\leq \eps_1\leq \theta_1$.

Intuitively, this suggests that $Y_2$ is a weaker channel.
Nevertheless, observe that this channel is \emph{not} degraded in the sense defined in Section~\ref{subsec:Bdegdef} (see (\ref{eq:Bdegraded})). For a given state $S=s$, the broadcast channel $\bc(\cdot,\cdot|\cdot,s)$ is stochastically degraded. 
In particular, one can define the following random variables,
\begin{align}
&A_s\sim \text{Bernoulli}(\pi_{s}) \,,\; \text{where $\pi_{s}\triangleq \frac{\eps_{s}-\theta_{s}}{1-2\theta_{s}}$} \,,
\\
&\tY_2=Y_1+A_S \mod 2 \,.
\end{align}
Then, $\tY_2$ is distributed according to
$
\prob{\tY_2=y_2 |  X=x,S=s} = \dWbc(y_2|x,s) 
$, 
and $X\Cbar (Y_1,S)\Cbar \tY_2$ form a Markov chain. However, since $X$ and $A_S$ depend on the state, it is not necessarily true that  $(X,S)\Cbar Y_1\Cbar \tY_2$ form a Markov chain, and the BSBC with correlated noises could be non-degraded.



We have the following results. 
\subsection*{Random Parameter BSBC with Correlated Noises}
First, we consider the random parameter BSBC $\avbc^q$, 
with a memoryless state $S\sim\text{Bernoulli}(q)$, for a given $0\leq q\leq 1$.
 Define the binary entropy function $h(x)=-x\log x-(1-x)\log(1-x)$, for $x\in [0,1]$, with logarithm to base $2$. 
We show that the capacity region of the random parameter BSBC $\avbc^q$ with degraded message sets with causal SI is given by 
\begin{align}
\BCrp=\BICrp=\bigcup_{0\leq \beta\leq 1}
\left\{
\begin{array}{lrl}
(R_0,R_1) \,:\; & R_0 &\leq   1-h(\beta*\wdelta) \,, \\
								& R_1 &\leq   h(\beta*\sdelta)-h(\sdelta)
\end{array}
\right\}\,,
\label{eq:BSBCcorrBCrp}
\end{align}
where
\begin{align}
\sdelta=&(1-q)\theta_0+q(1-\theta_1) \,,\; \nonumber\\
\wdelta=&(1-q)\eps_0+q(1-\eps_1) \,.
\label{eq:swdelta}
\end{align}
The proof is given in Appendix~\ref{app:AVBSBC2P1}.
It can be seen in the achievability proof that the parameter $\beta$ is related to the distribution of  $U_1$, and thus the RHS of (\ref{eq:BSBCcorrBCrp}) can be thought  of as a union over Shannon strategies.

\begin{center}
\begin{figure}[hbt]
\centering
\includegraphics[scale=0.5,trim={2.3cm 0 0 0},clip]{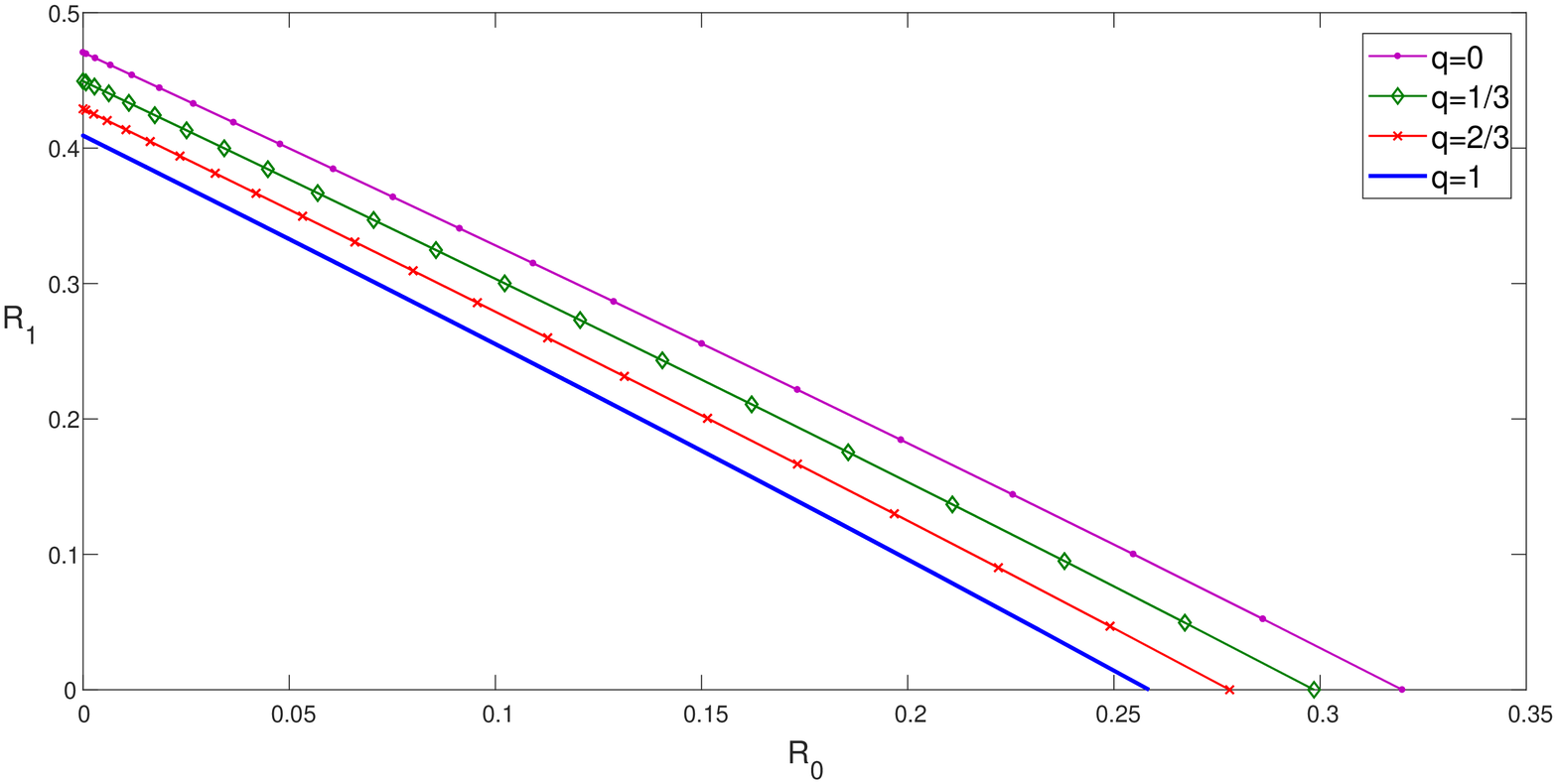}
\caption{The capacity region of the AVBC in Example~\ref{example:AVBSBC2}, the arbitrarily varying binary symmetric broadcast channel with correlated noises, with parameters that correspond to case 1. The area under the thick blue line is the  capacity region of the AVBC $\avbc$ with causal SI, with $\theta_0=0.12$, 
$\theta_1=0.85$, $\eps_0=0.18$ and $\eps_1=0.78$. 
The curves depict $\BICrp$ for $q=0,1/3,2/3,1$, where the capacity region of $\avbc$ is given by $\BCavc=\BrICav=\BICrp$ for $q=1$ 
(see (\ref{eq:BSBCcorrBCrp})). 
  }
\label{fig:BBSC2case1}
\end{figure}
\end{center}

\subsection*{Arbitrarily Varying BSBC with Correlated Noises}
We move to the arbitrarily varying BSBC with correlated noises.
As shown in Appendix~\ref{app:AVBSBC2P2},
the  capacity region of the arbitrarily varying BSBC $\Bavcig$ with degraded message sets without SI is given by
$
\BCavcig= \{(0,0)\} 
$. 
For the setting where causal SI is available at the encoder, we consider two cases.

\emph{Case 1:} Suppose that
$\theta_0\leq 1-\theta_1\leq \eps_0\leq 1-\eps_1\leq \frac{1}{2}$. That is, $S=1$ is a noisier channel state than $S=0$, for both users.
The  capacity region of the arbitrarily varying BSBC $\avbc$ with degraded message sets with causal SI is given by 
\begin{align}
\label{eq:Bex1Cavc2case1}
\BCavc= \BICrp\Big|_{q=1}= \bigcup_{0\leq \beta\leq 1}
\left\{
\begin{array}{lrl}
(R_0,R_1) \,:\; & R_0 &\leq   1-h(\beta*\eps_1) \,, \\
								& R_1 &\leq   h(\beta*\theta_1)-h(\theta_1)
\end{array}
\right\}\,.
\end{align}

It is shown in Appendix~\ref{app:AVBSBC2P2} that Condition $\sCond$ holds and $\BCavc=\BIRavc=\BrICav$.
Figure~\ref{fig:BBSC2case1} provides a graphical interpretation. 
The analysis shows that  Condition $\sCond$ implies that there exists $0\leq q^*\leq 1$ such that $\BCavc=\inC(\avbc^{q^*})$, where $\inC(\avbc^{q^*})\subseteq \inC(\Brp) $ for every $0\leq q\leq 1$. Indeed,
looking at Figure~\ref{fig:BBSC2case1}, it appears that the regions $\inC(\avbc^{q})$, for $0\leq q\leq 1$, form a well ordered set, hence 
 $\BCavc=\inC(\avbc^{q^*})$ with $q^*=1$.

\begin{center}
\begin{figure}[ht!]
\caption{The inner and outer bounds on the capacity region of the AVBC in Example~\ref{example:AVBSBC2}, the arbitrarily varying binary symmetric broadcast channel with correlated noises, with parameters that correspond to case 2, namely, $\theta_0=0.12$, $\theta_1=0.85$, $\eps_0=0.22$ and $\eps_1=0.88$. 
  }	
 \includegraphics[scale=0.51,trim={2.5cm 0 0 0},clip]{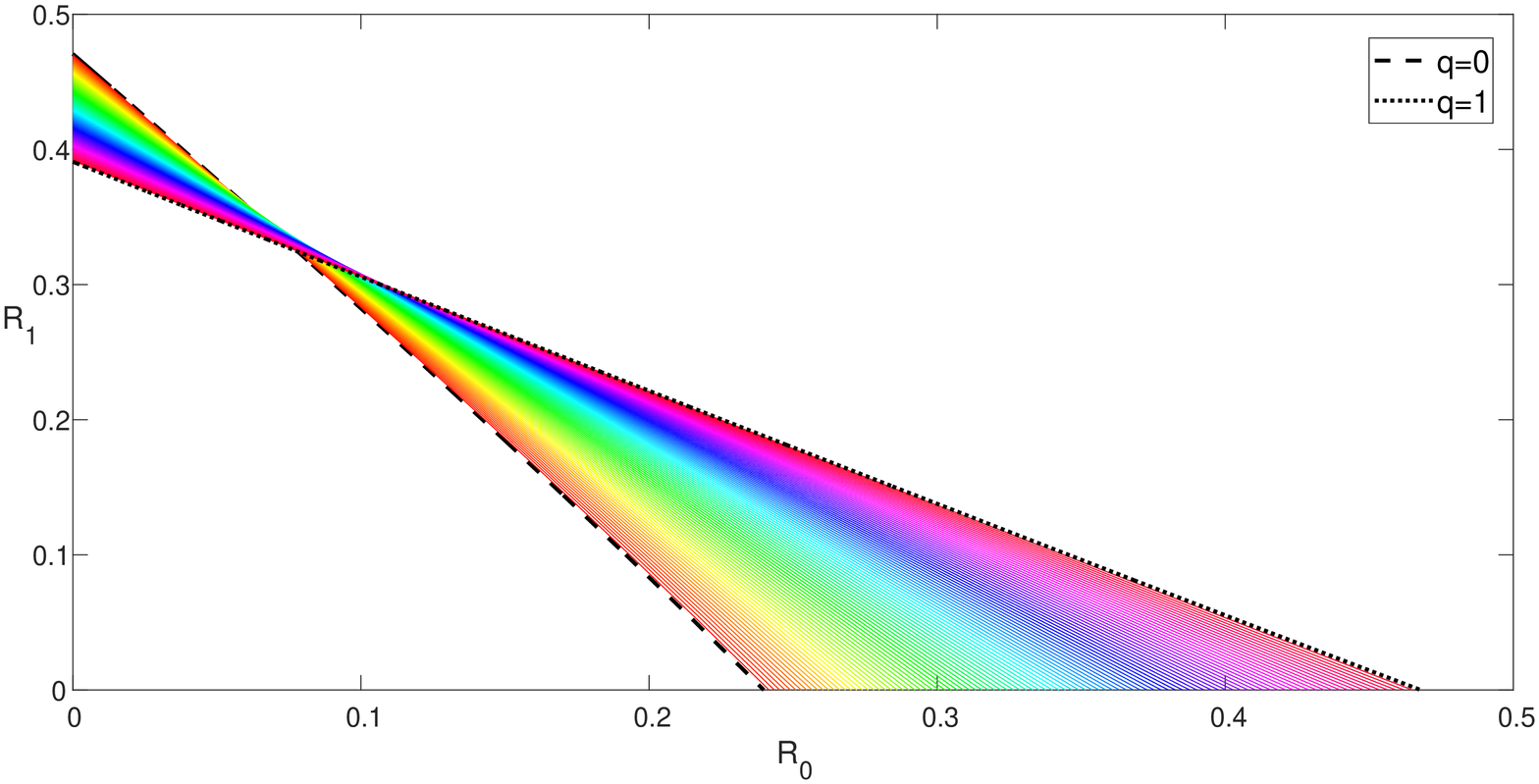}
\\
(a) 
The dashed and dotted lines depict the boundaries of $\inC(\avbc^{q=0})$ and $\inC(\avbc^{q=1})$, respectively. 
The colored lines depict $\inC(\avbc^{q})$ for a range of values of $0<q<1$. 
\\ 
\includegraphics[scale=0.51,trim={2.5cm 0 0 0},clip]{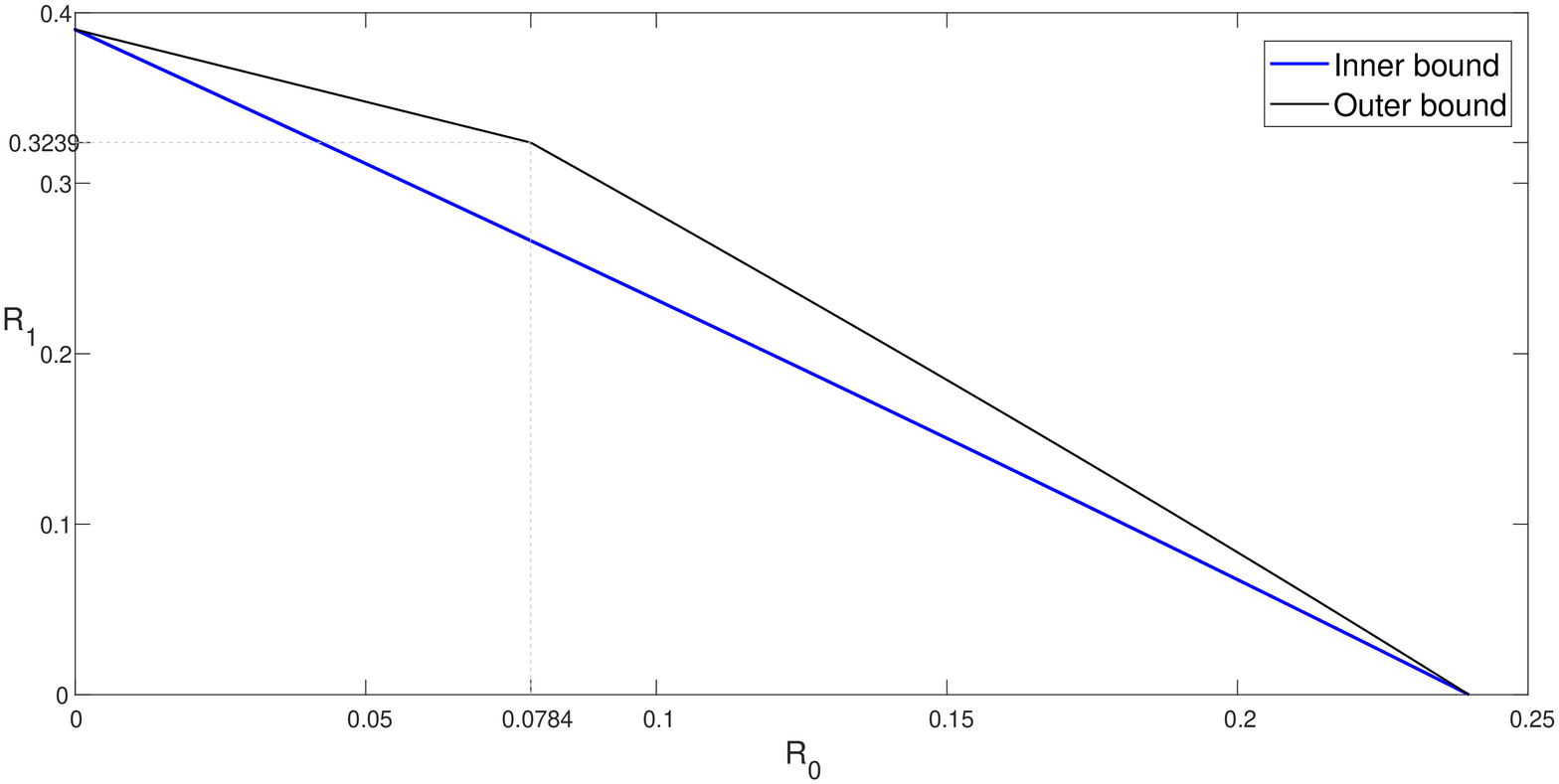}
\\
(b) The area under the thick blue line is the inner bound $\BIRavc$, and the area under the  thin line is the outer bound $\BrICav$. 
\label{fig:BBSC2case2}
\end{figure}

\end{center}

\emph{Case 2:} Suppose that
$\theta_0\leq 1-\theta_1\leq 1-\eps_1\leq\eps_0\leq  \frac{1}{2}$. That is, $S=1$ is a noisier channel state for user 1, whereas $S=0$ is noisier for user 2.
The  capacity region of the arbitrarily varying BSBC $\avbc$ with degraded message sets with causal SI is bounded by 
\begin{align}
\label{eq:Bex1Cavc2case2inner}
\BCavc\supseteq&
 \BIRavc= \bigcup_{0\leq \beta\leq 1}
\left\{
\begin{array}{lrl}
(R_0,R_1) \,:\; & R_0 &\leq   1-h(\beta*\eps_0) \,, \\
								& R_1 &\leq   h(\beta*\theta_1)-h(\theta_1)
\end{array}
\right\}\,,
\intertext{and}
\label{eq:Bex1Cavc2case2outer}
\BCavc\subseteq&
 \BrICav\subseteq \,\inC(\avbc^{q=0})\cap \inC(\avbc^{q=1}) \nonumber\\
=& \bigcup_{\substack{\; 0\leq \beta_0\leq 1 \,, \\  0\leq \beta_1\leq 1}}
\left\{
\begin{array}{lrl}
(R_0,R_1) \,:\; & R_0 &\leq   1-h(\beta_0*\eps_0) \,, \\
								& R_0 &\leq   1-h(\beta_1*\eps_1) \,, \\
								& R_1 &\leq   h(\beta_0*\theta_0)-h(\theta_0) \,,\\						
								& R_1 &\leq   h(\beta_1*\theta_1)-h(\theta_1)
\end{array}
\right\}
 \,.
\end{align}
The analysis is given in Appendix~\ref{app:AVBSBC2}. 
Figure~\ref{fig:BBSC2case2} provides a graphical interpretation. 
The dashed and dotted lines in Figure~\ref{fig:BBSC2case2}(a) depict the boundaries of $\inC(\avbc^{q=0})$ and $\inC(\avbc^{q=1})$, respectively. 
The colored lines depict $\inC(\avbc^{q})$ for a range of values of $0<q<1$. 
It appears that $\BrICav=\cap_{0\leq q\leq 1} \BICrp$ reduces to the intersection of the regions  $\inC(\avbc^{q=0})$ and $\inC(\avbc^{q=1})$. Figure~\ref{fig:BBSC2case2}(b) demonstrates the gap between the bounds in \mbox{case 2}. 

\end{example}
\begin{appendices}
\section{Proof of Lemma~\ref{lemm:BcompoundLowerB}}
\label{app:BcompoundLowerB}
We show that every rate pair $(R_0,R_1)\in\BIRcompound$ can be achieved using deterministic codes over the compound broadcast channel $\Bcompound$ with causal SI. 
We construct a code based on superposition coding with Shannon strategies, and decode using joint typicality  with respect to a channel state type, which is ``close" to some  $q\in\Qset$. 

We use the following notation. Basic method of types concepts are defined as in \cite[Chapter 2]{CsiszarKorner:82b}; including the definition of a type $\hP_{x^n}$ of a sequence $x^n$; a joint type $\hP_{x^n,y^n}$ and a conditional type $\hP_{x^n|y^n}$ of a pair of sequences $(x^n,y^n)$; and 
 a $\delta$-typical set $\tset(P_{X,Y})$ with respect to a distribution $P_{X,Y}(x,y)$. 
Define a set $\tQ$ of state types
\begin{align}
\tQ= \left\{ \emp_{s^n} \,:\; s^n\in\Aset^{\delta_1}(q) \,,\;\text{for some $q\in\Qset$}  \right\} \,,
\label{eq:BtQ}
\end{align}
where
\begin{align}
\label{eq:Bdelta1CompNoSI}
\delta_1\triangleq \frac{\delta}{2\cdot |\Sset|} \,,
\end{align}
where $\delta>0$ is arbitrarily small.
That is, $\tQ$ is the set of types that are $\delta_1$-close to some state distribution $q(s)$ in $\Qset$. 
Note that for any fixed $\delta$ (or $\delta_1$), for a sufficiently large $n$, the set $\tQ$ covers the set $\Qset$, and it is in fact a $\delta_1$-blowup of $\Qset$.
Now, a code for the compound broadcast channel with causal SI is constructed as follows.

\emph{Codebook Generation:} Fix the distribution $P_{U_0,U_1}$ 
 and the function $\encs(u_0,u_1,s)$. Generate $2^{nR_0}$ independent sequences at random,
\begin{align}
u_0^n(m_0) \sim  \prod_{i=1}^n P_{U_0}(u_{0,i}) \,,\;\text{for $m_0\in[1:2^{nR_0}]$}\,.
\end{align}
For every $m_0\in[1:2^{nR_0}]$, generate $2^{nR_1}$ sequences at random,  
 \begin{align}
u_1^n(m_0,m_1) \sim \prod_{i=1}^n P_{U_1|U_0}(u_{1,i}|u_{0,i}(m_0)) \,,\;\text{for $m_1\in[1:2^{nR_1}]$}\,,
\end{align}
conditionally independent given $u_0^n(m_0)$.

\emph{Encoding}: To send a pair of messages $(m_0,m_1)\in [1:2^{nR_0}]\times [1:2^{nR_1}]$, transmit at time $i\in[1:n]$,
\begin{align}
x_i=\encs\left( u_{0,i}(m_0),u_{1,i}(m_0,m_1),s_i \right) \,.
\end{align}

\emph{Decoding}: Let
 \begin{align}
\label{eq:BUchannelYL1causal}
P^{q}_{U_0,U_1,Y_1,Y_2}(u_0,u_1,y_1,y_2)=\sum_{s\in\Sset} q(s)
 P_{U_0,U_1}(u_0,u_1) 
 \bc\left( y_1, y_2| \encs(u_0,u_1,s),s \right) \,.
 \end{align}
Observing $y_2^n$, decoder 2 finds a unique $\tm_0\in[1:2^{nR_0}]$ such that
\begin{align}
 (u_0^n(\tm_0),y_2^n)\in\tset(P_{U_0} P^{q}_{Y_2|U_0}) \,,\quad\text{for some $q\in\tQ$} \,.
\end{align}  
If there is none, or more than one such $\tm_0\in[1:2^{nR_0}]$, then decoder 2 declares an error.

Observing $y_1^n$, decoder 1 finds a unique pair of messages $(\hm_0,\hm_1)\in [1:2^{nR_0}]\times [1:2^{nR_1}]$ such that
\begin{align}
(u_0^n(\hm_0),u_1^n(\hm_0,\hm_1),y_1^n)\in\tset( P_{U_0,U_1} P^q_{Y_1|U_0,U_1} )
 \,,\quad\text{for some $q\in\tQ$} \,.
\end{align}
If there is none, or more than such pair $(\hm_0,\hm_1)$, then decoder 1 declares an error.
We note that using the set of types $\tQ$ instead of the original set of state distributions $\Qset$ alleviates the analysis, since
 $\Qset$ is not necessarily finite nor countable.

\emph{Analysis of Probability of Error}:
Assume without loss of generality that the users sent the message pair
$(M_0,M_1)=(1,1)$. Let $q(s)\in\Qset$ denote the \emph{actual} state distribution chosen by the jammer.
By the union of events bound,
\begin{align}
\label{eq:BcompLerr12}
\err(q,\code)\leq \prob{\tM_0\neq 1}+\prob{(\hM_0,\hM_1)\neq (1,1)} \,,
\end{align}
where the conditioning on $(M_0,M_1)=(1,1)$ is omitted for convenience of notation.
The error event for decoder 2 is the union of the following events. 
\begin{align}
\Eset_{2,1} =&\{ (U_0^n(1),Y_2^n)\notin \tset(P_{U_0} P^{q'}_{Y_2|U_0}) \;\text{ for all $q'\in\tQ$} \} \,, \\
\Eset_{2,2} =&\{ (U_0^n(m_0),Y^n)\in \tset(P_{U_0} P^{q'}_{Y_2|U_0}) \;\text{ for some $m_0\neq 1,\, q'\in\tQ$} \} \,.
\end{align}
 Then, by the union of events bound,
\begin{align}
\label{eq:Bub2}
 \prob{\tM_2\neq 1}\leq \prob{\Eset_{2,1}}+ \prob{\Eset_{2,2}} \,.
\end{align}  
 Considering the first term, we claim that the event $\Eset_{2,1}$ implies that $(U_0^n(1),Y_2^n)\notin \Aset^{\nicefrac{\delta}{2}}(P_{U_0} P^{q''}_{Y_2|U_0})$ for all $q''\in\Qset$. 
	Assume to the contrary that $\Eset_{2,1}$ holds, but  there exists $q''\in\Qset$ such that 
	$(U_0^n(1),Y_2^n)\in \Aset^{\nicefrac{\delta}{2}}(P_{U_0} P^{q''}_{Y_2|U_0})$. 
	Then, for a sufficiently large $n$, there exists a type $q'(s)$ such that 
	$
	|q'(s)-q''(s)|\leq \delta_1 
	$ for all $s\in\Sset$. 
	It can then be inferred that $q'\in\tQ$ (see (\ref{eq:BtQ})), and 
	\begin{align}
	|P_{Y_2|U_0}^{q'}(y_2|u_0)-P_{Y_2|U_0}^{q''}(y_2|u_0)|\leq |\Sset|\cdot\delta_1=\frac{\delta}{2} \,,
	\end{align}
	for all $u_0\in\Uset_0$ and $y_2\in\Yset_2$ (see  (\ref{eq:Bdelta1CompNoSI}) and (\ref{eq:BUchannelYL1causal})).
	Hence,	$(U_0^n(1),Y_2^n)\in \Aset^{\delta}(P_{U_0} P^{q'}_{Y_2|U_0})$, which contradicts the first assumption. 
	Thus,
	\begin{align} 
	\label{eq:BllnRL}
	\prob{\Eset_{2,1}}\leq& \prob{(U_0^n(1),Y_2^n)\notin \Aset^{\nicefrac{\delta}{2}}(P_{U_0} P^{q''}_{Y_2|U_0}) \;\text{ for all $q''\in\Qset$} } \nonumber \\
	\leq& \prob{(U_0^n(1),Y_2^n)\notin \Aset^{\nicefrac{\delta}{2}}(P_{U_0} P^{q}_{Y_2|U_0})  } \,.
	\end{align}
The last expression  tends to zero exponentially as $n\rightarrow\infty$ by the law of large numbers and Chernoff's bound.
	
	Moving to the second term in the RHS of (\ref{eq:Bub2}), 
	we use the classic method of types considerations to bound $\prob{\Eset_{2,2}}$. 
	By the union of events bound 
and the fact that the number of type classes in $\Sset^n$ is bounded by $(n+1)^{|\Sset|}$, 
 we have that  
\begin{align}
&\prob{\Eset_{2,2}} \nonumber\\
\leq& (n+1)^{|\Sset|}
\cdot \sup_{q'\in\tQ} \prob{
(U_0^n(m_0),Y_2^n)\in \tset(P_{U_0} P^{q'}_{Y_2|U_0}) \;\text{ for some $m_0\neq 1$} 
}\,.
\label{eq:BE2poly}
\end{align}
For every $m_0\neq 1$,
\begin{align}
\prob{
(U_0^n(m_0),Y_2^n)\in \tset(P_{U_0} P^{q'}_{Y_2|U_0}) 
}
=&  \sum_{u_0^n\in\Uset_0^n} P_{U_0^n}(u_0^n) \cdot \prob{(u_0^n,Y_2^n)\in\tset(P_{U_0} P^{q'}_{Y_2|U_0})} 													\nonumber\\
=&  \sum_{u_0^n\in\Uset_0^n} P_{U_0^n}(u_0^n) \cdot \sum_{y_2^n \,:\; (u_0^n,y_2^n)\in \tset(P_{U_0} P^{q'}_{Y_2|U_0})} P_{Y_2^n}^q(y_2^n) \,,
\label{eq:BE2bound0} 
\end{align}
where the last equality holds since $U_0^n(m_0)$ is independent of $Y_2^n$ for every $m_0\neq 1$. Let 
 $(u_0^n,y_2^n)\in \tset(P_{U_0} P^{q'}_{Y_2|U_0})$. Then, $\,y_2^n\in\Aset^{\delta_2}(P_{Y_2}^{q'})$ with $\delta_2\triangleq |\Uset_0|\cdot\delta$. By Lemmas 2.6 and 2.7 in
 \cite{CsiszarKorner:82b},
\begin{align}
\label{eq:BpYbound}
P_{Y_2^n}^q(y_2^n)=2^{-n\left(  H(\hP_{y_2^n})+D(\hP_{y_2^n}||P_{Y_2}^q)
\right)}\leq 2^{-n H(\hP_{y_2^n})}
\leq 2^{-n\left( H_{q'}(Y_2) -\eps_1(\delta) \right)} \,,
\end{align}
where $\eps_1(\delta)\rightarrow 0$ as $\delta\rightarrow 0$. Therefore, by (\ref{eq:BE2poly})$-$(\ref{eq:BpYbound}),
\begin{align}
& \prob{\Eset_{2,2}}           																										\nonumber\\
\leq& (n+1)^{|\Sset|}																															\nonumber\\
& \cdot \sup_{q'\in\tQ} \left[
2^{nR_0} \cdot \sum_{u_0^n\in\Uset_0^n} P_{U_0^n}(u_0^n)\cdot |\{y_2^n\,:\; (u_0^n,y_2^n)\in\tset(P_{U_0} P_{Y_2|U_0}^{q'})\}| \cdot 
 2^{-n\left( H_{q'}(Y_2) -\eps_1(\delta) \right)}		 \right]	  												\nonumber\\
\leq& (n+1)^{|\Sset|}\cdot \sup_{q'\in\Qset} 
2^{-n[ I_{q'}(U_0;Y_2) 
-R_0-\eps_2(\delta) ]} \label{eq:BexpCR2} \,,
\end{align}
with $\eps_2(\delta)\rightarrow 0$ as $\delta\rightarrow 0$, 
where the last inequality is due to \cite[Lemma 2.13]{CsiszarKorner:82b}. The RHS of (\ref{eq:BexpCR2})
  tends to zero exponentially as $n\rightarrow\infty$, provided that $R_0<\inf_{q'\in\Qset} I_{q'}(U_0;Y_2)
-\eps_2(\delta)$.  

Now, consider the error event of decoder 1. For every $(m_0,m_1)\in [1:2^{nR_0}]\times [1:2^{nR_1}]$, define the event
\begin{align}
&\Eset_{1,1}(m_0,m_1)=\{ (U_0^n(m_0),U_1^n(m_0,m_1),Y_1^n)\in\tset(P_{U_0,U_1} P^{q'}_{Y_1|U_0,U_1})  \,,\;\text{for some $q'\in\tQ$}\}\,.
\end{align}
Then, the error event is bounded by
\begin{align}
&\left\{(\hM_0,\hM_1)\neq (1,1)\right\}
\subseteq \,\Eset_{1,1}(1,1)^c \,\cup\; \bigcup_{m_1\neq 1}\Eset_{1,1}(m_1,1)
\,\cup\; \bigcup_{ \substack{m_1\in [1:2^{nR_1}] \,,\\
 m_0\neq 1}
}\Eset_{1,1}(m_0,m_1)																																
  \,.
\end{align}
Thus, by the union of events bound,
\begin{align}
&\prob{(\hM_0,\hM_1)\neq (1,1)} \nonumber\\
\leq&\prob{ \Eset_{1,1}(1,1)^c} + \sum_{\substack{ m_1\in [1:2^{nR_1}] \,,\\ 
 m_0\neq 1}
} \prob{\Eset_{1,1}(m_0,m_1) }
+ \sum_{m_1\neq 1} \prob{\Eset_{1,1}(m_1,1) } \nonumber\\
\leq&
2^{-\theta n}+2^{-n\left(\inf\limits_{q'\in\Qset} I_{q'}(U_0,U_1;Y_1)-R_0-R_1-\eps_3(\delta) \right)}+\sum_{m_1\neq 1}\prob{\Eset_{1,1}(m_1,1) }
\,,
\label{eq:Blowerexp1}
\end{align}
where the last inequality follows from the law of large numbers and type class considerations used before, with $\eps_3(\delta)\rightarrow 0$ as $\delta\rightarrow 0$.  
The middle term in the RHS of (\ref{eq:Blowerexp1}) exponentially tends to zero as $n\rightarrow\infty$ provided that $R_0+R_1<\inf\limits_{q'\in\Qset} I_{q'}(U_0,U_1;Y_1)-\eps_3(\delta)$.
 It remains for us to bound the last sum. Using similar type class considerations, we have that for every $q'\in\tQ$ and $m_1\neq 1$,
\begin{align}
&\prob{ (  U_0^n(1),U_1^n(m_1,1),Y_1^n)\in \tset(P_{U_0,U_1} P^{q'}_{Y_1|U_0,U_1})}  \nonumber\\
=& \sum_{(u_0^n,u_1^n,y_1^n)\in \tset(P_{U_0,U_1} P^{q'}_{Y_1|U_0,U_1})} 
P_{U_0^n}(u_0^n)\cdot    P_{U_1^n|U_0^n}(u_1^n|u_0^n)\cdot  P^q_{Y_1^n|U_0^n}(y_1^n|u_0^n)\nonumber\\
\leq& 2^{n(H_{q'}(U_0,U_1,Y_1)+\eps_4(\delta))}\cdot 2^{-n(H(U_0)-\eps_4(\delta))}\cdot 2^{-n(H(U_1|U_0)-\eps_4(\delta))}\cdot 2^{-n(H_{q'}(Y_1|U_0)-\eps_4(\delta))}														\nonumber\\
=& 2^{-n(I_{q'}(U_1;Y_1|U_0)-4 \eps_4(\delta))} \,,
\label{eq:BI1}
\end{align}
where $\eps_4(\delta)\rightarrow 0$ as $\delta\rightarrow 0$. 
Therefore, the sum term in the RHS of (\ref{eq:Blowerexp1}) is bounded by 
\begin{align}
&\sum_{m_1\neq 1}\prob{\Eset_{1,1}(m_1,1) } \nonumber\\
=&\sum_{m_1\neq 1} \prob{
(U_0^n(1),U_1^n(m_1,1),Y_1^n)\in\tset(P_{U_0,U_1} P^{q'}_{Y_1|U_0,U_1})  \,,\;\text{for some $q'\in\tQ$}\}
}\nonumber\\
\leq& (n+1)^{|\Sset|} \cdot 2^{-n\left(\inf\limits_{q'\in\Qset} I_{q'}(U_1;Y_1|U_0)-R_1- \eps_5(\delta) \right)} \,,
\label{eq:BexpCR1}
\end{align}
where the last line follows from (\ref{eq:BI1}), and $\eps_5(\delta)\rightarrow 0$ as $\delta\rightarrow 0$. The last expression  tends to zero  exponentially as $n\rightarrow\infty$ and $\delta\rightarrow 0$ provided that $R_1<\inf_{q'\in\Qset} I_{q'}(U_1;Y_1|U_0)-\eps_5(\delta)$.  

The probability of error, averaged over the class of the codebooks, exponentially decays to zero  as $n\rightarrow\infty$. Therefore, there must exist a $(2^{nR_0},2^{nR_1},n,e^{-an})$ deterministic code, for a sufficiently large $n$.
\qed

\section{Proof of Theorem~\ref{theo:BcvC}}            
\label{app:BcvC}   
\section*{Part 1}
At the first part of the theorem it is assumed that the interior of the capacity region is non-empty, \ie $\interior{\BCcompound}\neq\emptyset$. 
 
 \begin{proof}[Achievability proof]
 We show that every rate pair $(R_0,R_1)\in\BICcompound$ can be achieved using a code based on Shannon strategies with the addition of a codeword \emph{suffix}. 
 At time $i=n+1$, having completed the transmission of the messages, the type of the state sequence $s^n$ is known to the encoder.
Following the assumption that the interior of the capacity region is non-empty,  the type of $s^n$  can be reliably communicated to both receivers 
 as a suffix, while the blocklength is increased by  $\nu>0$ additional channel uses, where $\nu$ is small compared to $n$. 
The receivers  first estimate the type of $s^n$, and then use joint typicality  with respect to the estimated type. 
  The details are provided below.  
	
Following the assumption that $\interior{\BCcompound}\neq\emptyset$, 
we have that for every $\eps_1>0$ and sufficiently large blocklength $\nu$, there exists a 
$(2^{\nu \tR_0},$ $2^{\nu \tR_1},\nu,\eps_1)$ code $\tcode=(\tf^\nu,\tg_1,\tg_2)$
 for the transmission of a type  $\hP_{s^n}$ at  positive rates $\tR_0>0$ and $\tR_1>0$. 
 Since the total number of types is polynomial in $n$ (see \cite{CsiszarKorner:82b}),  the type $\hP_{s^n}$ can be transmitted at a negligible rate, with a blocklength that grows a lot slower than $n$, \ie   
\begin{align}
\nu=o(n) \,.
\end{align}
We now construct a code $\code$ over the compound broadcast channel with causal SI, such that the blocklength is $n+o(n)$, and the rate 
$R'_n$ approaches $R$ as $n\rightarrow\infty$. 

\emph{Codebook Generation:} Fix the distribution $P_{U_0,U_1}$ 
 and the function $\encs(u_0,u_1,s)$. Generate $2^{nR_0}$ independent sequences $u_0^n(m_0)$, $m_0\in[1:2^{nR_0}]$, at random, each according to
 $\prod_{i=1}^n P_{U_0}(u_{0,i})$.
For every $m_0\in [1:2^{nR_0}]$, generate $2^{nR_1}$ sequences at random, 
\begin{align}
u_1^n(m_0,m_1) \sim \prod_{i=1}^n P_{U_1|U_0}(u_{1,i}|u_{0,i}(m_0)) \,,\;\text{for $m_1\in [1:2^{nR_1}]$}\,,
\end{align}
conditionally independent given $u_0^n(m_0)$.  
 Reveal the codebook of the message pair $(m_0,m_1)$ and the codebook of the type  $\hP_{s^n}$  to the encoder and the decoders.

\emph{Encoding}: To send a message pair $(m_0,m_1)\in [1:2^{nR_0}]\times [1:2^{nR_1}]$, transmit at time $i\in[1:n]$,
\begin{align}
x_i=\encs\left( u_{0,i}(m_0),u_{1,i}(m_0,m_1),s_i\right) \,.
\end{align}
At time $i\in [n+1:n+\nu]$, knowing the sequence of previous states $s^n$,  transmit 
\begin{align}
x_i=\tf_i(\hP_{s^n},\, s_{n+1},\ldots,s_{n+i} ) \,,
\end{align}
where $\hP_{s^n}$ is the type of the sequence $(s_1,\ldots,s_n)$. 
That is, the encoded type $\hP_{s^n}$ is transmitted as a suffix of the codeword.
We note that the type of the sequence $(s_{n+1},\ldots,s_{n+i})$ is not necessarily $\hP_{s^n}$, and it 
 is irrelevant for that matter, since the assumption that $\interior{\BCcompound}\neq\emptyset$ implies that 
there exists a  $(2^{\nu \tR_0},2^{\nu \tR_1},\nu,\eps_1)$ code $\tcode=(\tf^\nu,\tg_1,\tg_2)$ for the transmission of $\hP_{s^n}$, 
 with $\tR_0>0$ and $\tR_1>0$.

\emph{Decoding}: Let
 \begin{align}
\label{eq:BUchannelYL1causalUp}
P^{q}_{U_0,U_1,Y_1,Y_2}(u_0,u_1,y_1,y_2)=\sum_{s\in\Sset} q(s)
 P_{U_0,U_1}(u_0,u_1) 
 \bc\left( y_1, y_2| \encs(u_0,u_1,s),s \right) \,.
 \end{align}
 Decoder 2 receives the output sequence $y_2^{n+\nu}$. As a pre-decoding step, the receiver decodes the last $\nu$ output symbols, and finds an estimate of the type of the state sequence, 
$
\hq_2=\tg_2(y_{2,n+1},\ldots,y_{2,n+\nu}) 
$. 
 Then, given the output sequence $y_2^n$,  decoder 2 finds a unique $\tm_0\in[1:2^{nR_0}]$ such that
\begin{align}
 (u_0^n(\tm_0),y_2^n)\in\tset(P_{U_0} P^{\hq_2}_{Y_2|U_0}) \,.
\end{align}  
If there is none, or more than one such $\tm_0\in[1:2^{nR_0}]$, then decoder 2 declares an error.

Similarly, decoder 1 receives  $y_1^{n+\nu}$ and begins with decoding the type of the state sequence,
$
\hq_1=\tg_1(y_{1,n+1},\ldots,y_{1,n+\nu}) 
$. 
Then, decoder 1 finds a unique pair of messages $(\hm_0,\hm_1)\in [1:2^{nR_0}]\times [1:2^{nR_1}]$ such that 
\begin{align}
(u_0^n(\hm_0),u_1^n(\hm_0,\hm_1),y_1^n)\in \tset(P_{U_0,U_1} P^{\hq_1}_{Y_1|U_0,U_1}) \,.
\end{align}
If there is none, or more than one such pair  $(\hm_0,\hm_1)\in [1:2^{nR_0}]\times [1:2^{nR_1}]$, then decoder 1 declares an error.

\emph{Analysis of Probability of Error}:
By symmetry, 
we may assume without loss of generality that the users sent $(M_0,M_1)=(1,1)$. 
Let $q(s)\in\Qset$ denote the actual state distribution chosen by the jammer, and let $\qn(s^n)=\prod_{i=1}^n q(s_i)$. 
Then, by the union of events bound, the probability of error is bounded by 
\begin{align}
\err(q,\code)\leq \prob{\tM_0 \neq 1} + \prob{(\hM_0,\hM_1)\neq (1,1)} \,,
\end{align}
where the conditioning on $(M_0,M_1)=(1,1)$ is omitted for convenience of notation. 

Define the  events  
\begin{align}
&\Eset_{1,0} =\{ \hq_1\neq \hP_{S^n} \} 						 																		\label{eq:BE11}\\
&\Eset_{1,1}(m_0,m_1,q')=\{ (U_0^n(m_0),U_1^n(m_0,m_1),Y_1^n)\in \tset(P_{U_0,U_1} P^{q'}_{Y_1|U_0,U_1}) \}   \label{eq:BE12}\\
\intertext{ 
and
 }
&\Eset_{2,0} =\{ \hq_2\neq \hP_{S^n} \} 																								\label{eq:BE21}\\
&\Eset_{2,1}(m_0,q') =\{ (U_0^n(m_0),Y_2^n)\in \tset(P_{U_0} P^{q'}_{Y_2|U_0})\} \,,						\label{eq:BE22} 
\end{align}
for every $m_0\in [1:2^{nR_0}]$, $m_1\in [1:2^{nR_1}]$, 
 and $q'\in\pSpace(\Sset)$. 
The error event of decoder 2 is bounded by
\begin{align}
\left\{ \tM_2\neq 1 \right\}\subseteq\,&\Eset_{2,0}\cup\Eset_{2,1}(1,\hq_2\,)^c\cup
\bigcup_{m_0\neq 1} \Eset_{2,1}(m_0,\hq_2\,) \nonumber\\
=\,& 
\Eset_{2,0} \,\cup\,
\left(  \Eset_{2,0}^c\cap\Eset_{2,1}(1,\hq_2\,)^c  \right) \,\cup\,
\left( \bigcup_{m_0\neq 1} \Eset_{2,0}^c\cap\Eset_{2,1}(m_0,\hq_2\,) \right) 
 \,. \nonumber
\end{align}
 By the union of events bound,
\begin{align}
&\prob{\tM_2\neq 1}																																	\nonumber\\	
&\leq 
 \prob{\Eset_{2,0}} +
\prob{\Eset_{2,0}^c\cap\Eset_{2,1}(1,\hq_2\,)^c }                             
+  \prob{\bigcup_{m_0\neq 1} \Eset_{2,0}^c\cap  \Eset_{2,1}(m_0,\hq_2\,)} \,.
\label{eq:BcvcErr1}
\end{align}

 Since the code $\tcode$ for the transmission of the type is a $(2^{\nu \tR_0},2^{\nu \tR_1},\nu,\eps_1)$ code,
where $\eps_1>0$ is arbitrarily small, we have that the probability of erroneous decoding of the type is bounded by
 \begin{align}
\label{eq:decq}
\prob{\Eset_{1,0}\cup \Eset_{2,0}}\leq\eps_1 \,.
\end{align}
 Thus, 
the first term in the RHS of (\ref{eq:BcvcErr1}) is bounded by $\eps_1$. Then, we maniplute the last two terms as follows.  
\begin{align}
\prob{\tM_2\neq 1}
\leq& \;
\sum_{s^n\in\Aset^{\delta_2}(q)}\qn(s^n)
\cprob{\Eset_{2,0}^c\cap\Eset_{2,1}(1,\hq_2\,)^c  }{S^n=s^n}
																																			\nonumber\\
&+
\sum_{s^n\notin\Aset^{\delta_2}(q)} \qn(s^n)
\cprob{\Eset_{2,0}^c\cap\Eset_{2,1}(1,\hq_2\,)^c }{S^n=s^n}
																																			\nonumber\\
&+
\sum_{s^n\in\Aset^{\delta_2}(q)} \qn(s^n)
\cprob{\bigcup_{m_0\neq 1} \Eset_{2,0}^c\cap  \Eset_{2,1}(m_0,\hq_2\,)}{S^n=s^n}
																																			\nonumber\\
&+
\sum_{s^n\notin\Aset^{\delta_2}(q)} \qn(s^n)
\cprob{\bigcup_{m_0\neq 1} \Eset_{2,0}^c\cap  \Eset_{2,1}(m_0,\hq_2\,)}{S^n=s^n}  +\eps_1			\,,	
\label{eq:BEcompound}																														
\end{align}
where 
\begin{align}
\delta_2 \triangleq\frac{1}{2|\Sset|
}\cdot\delta 	\,.		\label{eq:Bdelta1}
\end{align}
 Next we show that the first and the third sums in (\ref{eq:BEcompound}) 
 tend to zero as $n\rightarrow\infty$. 

Consider a given $s^n\in\Aset^{\delta_2}(q)$. For notational convenience, denote
 \begin{align}
\label{eq:BempC}
q''=\hP_{s^n} \,.
\end{align} 
 Then, by the definition of the $\delta$-typical set, we have that
$
|q''(s)-q(s)|\leq\delta_2 
$ 
for all $s\in\Sset$. 
It follows that
\begin{align}
&|P_{U_0}(u_0)P_{Y_2|U_0}^{q''}(y|u_0)-P_{U_0}(u_0)P_{Y_2|U_0}^q(y_2|u_0)|\nonumber\\
\leq&  \delta_2\cdot\sum_{s,u_1} P_{U_1|U_0}
(u_1|u_0) \wbc(y_2|\encs(u_0,u_1,s),s)
\leq 
 \delta_2\cdot|\Sset|=\frac{\delta}{2} \,,
\label{eq:BpUYclose}
\end{align}
for all $u_0\in\Uset_0$ and $y_2\in\Yset_2$, where the last equality follows from (\ref{eq:Bdelta1}).

Consider the first sum in the RHS of (\ref{eq:BEcompound}). 
Given a state sequence  $s^n\in\Aset^{\delta_2}(q)$,  we have that
\begin{align}
&\cprob{\Eset_{2,0}^c\cap\Eset_{2,1}(1,\hq_2\,)^c }{S^n=s^n}    = 
\cprob{\Eset_{2,0}^c\cap\Eset_{2,1}(1,\hP_{s^n}\,)^c }{S^n=s^n} \nonumber\\ =&
\cprob{\Eset_{2,0}^c\cap\Eset_{2,1}(1,q''\,)^c }{S^n=s^n}       \nonumber\\ =&
\cprob{\Eset_{2,0}^c }{\Eset_{2,1}(1,q'')^c, S^n=s^n}\cdot\cprob{\Eset_{2,1}(1,q'')^c\,) }{S^n=s^n} 
 \,, \label{eq:BEsum1}
\end{align}
where the first equality follows from (\ref{eq:BE21}),  
 and the second equality follows from (\ref{eq:BempC}). 
Then,
\begin{align}
\cprob{\Eset_{2,0}^c\cap\Eset_{2,1}(1,\hq_2\,)^c }{S^n=s^n}
\leq& \cprob{\Eset_{2,1}(1,q'')^c\, }{S^n=s^n} 																\nonumber\\
=& \prob{\, (U_0^n(1),Y_2^n)\notin\Aset^{\delta}(P_{U_0} P^{q''}_{Y_2|U_0})  \,\big|\; S^n=s^n  }
\,.
\label{eq:BE2bound1} 
\end{align}
 Now, suppose that $(U_0^n(1),Y_2^n)\in\Aset^{\nicefrac{\delta}{2}}(P_{U_0} P^q_{Y_2|U_0})$, where $q$ is the actual state distribution. By (\ref{eq:BpUYclose}), in this case we have that $(U_0^n(1),Y_2^n)\in\tset(P_{U_0} P^{q''}_{Y_2|U_0})$. 
 Hence,  (\ref{eq:BE2bound1}) implies that
\begin{align}
\label{eq:BE2causal}
&\cprob{\Eset_{2,0}^c\cap\Eset_{2,1}(1,\hq_2\,)^c }{S^n=s^n} 
\leq
\prob{\, (U_0^n(1),Y_2^n)\notin\Aset^{\nicefrac{\delta}{2}}(P_{U_0} P^q_{Y_2|U_0})  \,\big|\; S^n=s^n  }
\,.
\end{align}
The first sum in the RHS of (\ref{eq:BEcompound}) 
 is then bounded as follows. 
\begin{align}
&\sum_{s^n\in\Aset^{\delta_2}(q)} \qn(s^n)\cprob{\Eset_{2,0}^c\cap\Eset_{2,1}(1,\hq_2\,)^c }{S^n=s^n} 											\nonumber\\
&\leq
\sum_{s^n\in\Aset^{\delta_2}(q)} \qn(s^n) \prob{\, (U_0^n(1),Y_2^n)\notin\Aset^{\nicefrac{\delta}{2}}(P_{U_0} P^q_{Y_2|U_0})  \,\big|\; S^n=s^n } 				\nonumber\\
&\leq
\sum_{s^n\in\Sset^n} \qn(s^n) \prob{\, (U_0^n(1),Y_2^n)\notin\Aset^{\nicefrac{\delta}{2}}(P_{U_0} P^q_{Y_2|U_0})  \,\big|\; S^n=s^n } 										\nonumber\\
&=
\prob{\, (U_0^n(1),Y_2^n)\notin\Aset^{\nicefrac{\delta}{2}}(P_{U_0} P^q_{Y_2|U_0})  }\leq \eps_2 \,,
\label{eq:2sum1}
\end{align}
for a sufficiently large $n$, where the last inequality follows from  the law of large numbers.

We bound the third sum in the RHS of (\ref{eq:BEcompound}) 
 using similar arguments. If $(U_0^n(m_0),Y_2^n)\in\tset(P_{U_0} P_{Y_2|U_0}^{q''})$, then
 $(U_0^n(m_0),Y_2^n)\in\Aset^{\nicefrac{3\delta}{2}}(P_{U_0} P_{Y_2|U_0}^q)$, due to (\ref{eq:BpUYclose}). Thus, for every $s^n\in\Aset^{\delta_2}(q)$,
\begin{align}
\cprob{\bigcup_{m_0\neq 1} \Eset_{2,0}^c\cap\Eset_{2,1}(m_0,\hq_2\,) }{S^n=s^n}
\leq&  \sum_{m_0\neq 1} \cprob{\Eset_{2,1}(m_0,q'')}{S^n=s^n} \nonumber\\
=& \sum_{m_0\neq 1} \prob{\, (U_0^n(m_0),Y_2^n)\in\tset(P_{U_0} P^{q''}_{Y_2|U_0}) 
\,\big|\; S^n=s^n }
\nonumber\\
\leq& \sum_{m_0\neq 1} \prob{\, (U_0^n(m_0),Y_2^n)\in\Aset^{\nicefrac{3\delta}{2}}(P_{U_0} P^q_{Y_2|U_0}) 
  \,\big|\; S^n=s^n } \,.
\end{align}
This, in turn, implies that the third sum in the RHS of (\ref{eq:BEcompound}) 
is bounded by
\begin{align}
&\sum_{s^n\in\Aset^{\delta_2}(q)} \qn(s^n)\cprob{\bigcup_{m_0\neq 1} \Eset_{2,0}^c\cap\Eset_{2,1}(m_0,\hq\,) }{S^n=s^n} 		\nonumber\\
\leq&
\sum_{s^n\in\Sset^n}\sum_{m_0\neq 1} \qn(s^n)\cdot \prob{\, (U_0^n(m_0),Y_2^n)\in\Aset^{\nicefrac{3\delta}{2}}(P_{U_0} P^q_{Y_2|U_0}) 
 \,\big|\; S^n=s^n }\nonumber\\
=&
\sum_{m_0\neq 1} \prob{\, (U_0^n(m_0),Y_2^n)\in\Aset^{\nicefrac{3\delta}{2}}(P_{U_0} P^q_{Y_2|U_0}) \, }\nonumber\\
\leq&   2^{-n[ I_{q}(U_0;Y_2)-R_0-\eps_2(\delta) ]} \label{eq:2sum2} \,, 
\end{align} 
with $\eps_2(\delta)\rightarrow 0$ as $\delta\rightarrow 0$. The last inequality follows from standard type class considerations. The RHS of  (\ref{eq:2sum2})  tends to zero as $n\rightarrow\infty$, provided that 
$
R_0< I_{q}(U_0;Y_2)-\eps_2(\delta) 
$. 
Then, it follows from the law of large numbers that the second and fourth sums in the RHS of  (\ref{eq:BEcompound})
tend to zero as $n\rightarrow\infty$. Thus, by 
 (\ref{eq:2sum1}) and (\ref{eq:2sum2}), we have that 
the probability of error of decoder 2, $\prob{\tM_2\neq 1}$, tends to zero as $n\rightarrow\infty$.

Now, consider the error event of decoder 1,
\begin{align}
\left\{(\hM_0,\hM_1)\neq (1,1)\right\} 									
\subseteq&\,
\Eset_{1,0} \,\cup\;  \,\Eset_{1,1}(1,1,\hq_1)^c \,
\cup\; \bigcup_{ \substack{m_0\neq 1 \,,\\ m_1\in [1:2^{nR_1}] }
}\Eset_{1,1}(m_0,m_1,\hq_1)		
\cup\;
 \bigcup_{m_1\neq 1}\Eset_{1,1}(1,m_1,\hq_1)
  \,.
\end{align} 
Thus, by the union of events bound,
\begin{align}
\prob{(\hM_0,\hM_1)\neq (1,1)} 																									
&\leq \prob{\Eset_{1,0}}  +
\prob{ \Eset_{1,0}^c \cap \Eset_{1,1}(1,1,\hq_1)^c} +  \prob{\bigcup_{\substack{m_0\neq 1 \,,\\ m_1\in [1:2^{nR_1}]} } \Eset_{1,0}^c \cap \Eset_{1,1}(m_0,m_1,\hq_1) }																																		\nonumber\\
&+  \prob{\bigcup_{m_1\neq 1}  \Eset_{1,0}^c \cap \Eset_{1,1}(m_1,1,\hq_1) } 
\,.
\label{eq:Blowerexp1c}
\end{align}
By (\ref{eq:decq}), the first term is bounded by $\eps_1$, and 
as done above, we write
\begin{align}
\prob{(\hM_0,\hM_1)\neq (1,1)} 
\leq& 
\sum_{s^n\in\Aset^{\delta_2}(q)}\qn(s^n)
\cprob{ \Eset_{1,0}^c \cap \Eset_{1,1}(1,1,\hP_{s^n})^c}{S^n=s^n}			\nonumber\\
&+
\sum_{s^n\in\Aset^{\delta_2}(q)} \qn(s^n)
\cprob{\bigcup_{\substack{m_1\in [1:2^{nR_1}]\\ m_0\neq 1}}  \Eset_{1,0}^c \cap \Eset_{1,1}(m_0,m_1,\hP_{s^n})}{S^n=s^n}
\nonumber\\
&+
\sum_{s^n\in\Aset^{\delta_2}(q)} \qn(s^n)
\cprob{\bigcup_{m_1\neq 1}  \Eset_{1,0}^c \cap \Eset_{1,1}(m_1,1,\hP_{s^n})}{S^n=s^n}
\nonumber\\
& +3\cdot\prob{S^n\notin\Aset^{\delta_2}(q)}+\eps_1 \,,
\label{eq:BEcausal1}
\end{align}
where $\delta_2$ is given by (\ref{eq:Bdelta1}). 
 By the law of large numbers, the probability $\prob{S^n\notin\Aset^{\delta_2}(q)}$ tends to zero as $n\rightarrow\infty$.
As for the sums, we use similar arguments to those used above.

 We have that for a given $s^n\in\Aset^{\delta_2}(q)$,
\begin{align}
&|P_{U_0,U_1}(u_0,u_1) P^{q''}_{Y_1|U_0,U_1}(y_1|u_0,u_1)-P_{U_0,U_1}(u_0,u_1)P^{q}_{Y_1|U_0,U_1}(y_1|u_0,u_1)| \nonumber\\
&\leq 
\delta_2 \cdot \sum_{s\in\Sset} \sbc(y_1|\encs(u_0,u_1,s) 
\leq |\Sset|\cdot \delta_2=\frac{\delta}{2} \,,
\label{eq:Btset1}
\end{align}
with $q''=\hP_{s^n}$, where the last equality follows from (\ref{eq:Bdelta1}).

The first sum in the RHS of (\ref{eq:BEcausal1}) is bounded by 
\begin{align}
&\sum_{s^n\in\Aset^{\delta_2}(q)}\qn(s^n)
\cprob{ \Eset_{1,0}^c \cap \Eset_{1,1}(1,1,\hP_{s^n})^c}{S^n=s^n} \nonumber\\
&\leq \sum_{s^n\in\Sset^n}\qn(s^n)
\cprob{ (U_0^n(1),U_1^n(1,1),Y_1^n )\notin \Aset^{\nicefrac{\delta}{2}}(P_{U_0,U_1} P^{q}_{Y_1|U_0,U_1})  }{S^n=s^n}\nonumber\\
&=
\prob{ (U_0^n(1),U_1^n(1,1),Y_1^n )\notin \Aset^{\nicefrac{\delta}{2}}(P_{U_0,U_1} P^q_{Y_1|U_0,U_1})}\leq \eps_2\,.
\end{align}
 The last inequality follows from the law of large numbers, for a sufficiently large $n$.

The second sum in the RHS of (\ref{eq:BEcausal1}) is bounded by 
\begin{align}
\sum_{s^n\in\Aset^{\delta_2}(q)} \qn(s^n)
\cprob{\bigcup_{\substack{m_1\in [1:2^{nR_1}]\\ m_0\neq 1}}  \Eset_{1,0}^c \cap \Eset_{1,1}(m_0,m_1,\hP_{s^n})}{S^n=s^n} 
\leq 2^{-n( I_q(U_0,U_1;Y_1)-R_0-R_1-\eps_3(\delta)} \,,
\end{align}
with $\eps_3(\delta)\rightarrow 0$ as $n\rightarrow \infty$ and $\delta\rightarrow 0$. This is obtained following the same analysis as for decoder 2. Then, the second sum tends to zero provided that  
$
R_0+R_1 < I_q(U_0,U_1;Y_1)-\eps_3(\delta) 
$. 

The third sum in the RHS of (\ref{eq:BEcausal1}) is bounded by 
\begin{align}
\sum_{s^n\in\Aset^{\delta_2}(q)} \qn(s^n)
\cprob{\bigcup_{m_1\neq 1}  \Eset_{1,0}^c \cap \Eset_{1,1}(m_1,1,\hP_{s^n})}{S^n=s^n} 
\leq& 
\sum_{s^n\in\Aset^{\delta_2}(q)} \sum_{m_1\neq 1}  \qn(s^n)
\cprob{  \Eset_{1,1}(m_1,1,\hP_{s^n})}{S^n=s^n} \,.
\end{align}
For every $s^n\in\Aset^{\delta_2}(q)$, it follows from (\ref{eq:Btset1}) that the event $\Eset_{1,1}(m_1,1,\hP_{s^n})$
implies that 
$
(U_0^n(1),U_1^n(m_1,1),Y_1^n)\in$\\ $\Aset^{\nicefrac{3\delta}{2}}(P^q_{U_0,U_1,Y_1}) 
$. 
  Thus,  the sum is bounded by 
\begin{align}
&\sum_{s^n\in\Aset^{\delta_2}(q)} \qn(s^n)
\cprob{\bigcup_{m_1\neq 1}  \Eset_{1,0}^c \cap \Eset_{1,1}(m_1,1,\hP_{s^n})}{S^n=s^n} 
\leq
2^{-n(I_q(U_1;Y_1|U_0)-R_1-\delta_3)} \,,
\label{eq:boundU1gU0}
\end{align}
where $\delta_3\rightarrow 0$ as $\delta\rightarrow 0$. Then, the RHS of (\ref{eq:boundU1gU0}) tends to zero as $n\rightarrow\infty$ provided that $R_1<I_q(U_1;Y_1|U_0)-\delta_3$.

We conclude that the RHS of both (\ref{eq:BEcompound}) and (\ref{eq:BEcausal1}) tend to zero as $n\rightarrow \infty$. Thus, the overall probability of error, averaged over the class of the codebooks, 
 decays to zero  as $n\rightarrow\infty$. Therefore, there must exist a $(2^{nR_0},2^{nR_1},n,\eps)$ deterministic code, for a sufficiently large $n$. 
\end{proof}

\begin{proof}[Converse proof]
First, we claim that it can be assumed that $U_0 \Cbar U_1\Cbar X$ form a Markov chain.
Define the following region,
\begin{align}
\inR_{M,out}(\Bcompound)= \bigcap_{q(s)\in\Qset} 
\bigcup_{p(u_0,u_1),\,\widetilde{\encs}(u_1,s)}
\left\{
\begin{array}{lrl}
(R_0,R_1) \,:\; & R_0 	\leq&   I_q(U_0;Y_2) \,, \\
								& R_1 	\leq&   I_q(U_1;Y_1|U_0) \,\\
								&R_0+R_1\leq&   I_q(U_0,U_1;Y_1)	
\end{array}
\right\} \,,
\end{align}
subject to $X=\widetilde{\encs}(U_1,S)$. Clearly, $\inR_{M,out}(\Bcompound)\subseteq \BICcompound$, since $\inR_{M,out}(\Bcompound)$ is obtained by 
restriction of the function $\encs(u_0,u_1,s)$ in the union on the RHS of (\ref{eq:BcvCI}). Moreover, we have that $\inR_{M,out}(\Bcompound)\supseteq \BICcompound$, since, given some $U_0$, $U_1$ and $\encs(u_0,u_1,s)$, we can define a new strategy variable $\tU_1=(U_0,U_1)$, and then $X$ is a deterministic function of $(\tU_1,S)$.

As $\BICcompound=\inR_{M,out}(\Bcompound) $, it can now be assumed that 
$U_0 \Cbar U_1\Cbar X\Cbar (Y_1,Y_2)$ form a Markov chain, hence $I_q(U_0,U_1;Y_1)=I_q(U_1;Y_1)$. Then, by similar arguements to those used in \cite{KornerMarton:77p} (see also \cite[Chapter 16]{CsiszarKorner:82b}), we have that   
\begin{align}
\BICcompound= \bigcap_{q(s)\in\Qset} 
\bigcup_{p(u_0,u_1),\,\widetilde{\encs}(u_1,s)}
\left\{
\begin{array}{lrl}
(R_0,R_1) \,:\; & R_0 	\leq&   I_q(U_0;Y_2) \,, \\
								&R_0+R_1 	\leq&   I_q(U_1;Y_1|U_0)+I_q(U_0;Y_2) \,\\
								&R_0+R_1\leq&   I_q(U_1;Y_1)	
\end{array}
\right\} \,.
\label{eq:BcompConvEquiv2}
\end{align}
We show that for every sequence of $(2^{nR_0},2^{nR_1},n,\theta_n)$ codes, with $\lim_{n\rightarrow\infty}\theta_n=0$, we have that $(R_0,R_1)$ belongs to the set above.

Define the following random variables,
\begin{align}
U_{0,i}\triangleq (M_0,Y_1^{i-1},Y_{2,i+1}^n) \,,\; U_{1,i}\triangleq (M_0,M_1,S^{i-1}) \,.
\end{align}
It follows that $X_i$ is a deterministic function of $(U_{1,i},S_i)$, and 
since the state sequence is memoryless, we have that $S_i$ is independent of $(U_{0,i},U_{1,i})$. 
Next, by Fano's inquality,
\begin{align}
& nR_0\leq I_q(M_0;Y_2^n)+n\eps_n \,, \label{eq:convBcompound1}
\\
& n(R_0+R_1)\leq I_q(M_0,M_1;Y_1^n)+n\eps_n \,, \label{eq:convBcompound2}
\\
& n(R_0+R_1)\leq I_q(M_1;Y_1^n|M_0)+I_q(M_0;Y_2^n)+n\eps_n \,, \label{eq:convBcompound3}
\end{align}
where $\eps_n\rightarrow 0$ as $n\rightarrow\infty$. 
Applying the chain rule, we have that (\ref{eq:convBcompound1}) is bounded by 
\begin{align}
&I_q(M_0;Y_2^n)=\sum_{i=1}^n I_q(M_0;Y_{2,i} | Y_{2,i+1}^n)\leq \sum_{i=1}^n I_q(U_{0,i};Y_{2,i}) \,,
\label{eq:convBcompound4}
\end{align}
and (\ref{eq:convBcompound2}) is bounded by 
\begin{align}
I_q(M_0,M_1;Y_1^n)=& \sum_{i=1}^n I_q(M_0,M_1;Y_{1,i}|Y_1^{i-1}) 
\leq \sum_{i=1}^n I_q(U_{0,i},U_{1,i};Y_{1,i}) 
= \sum_{i=1}^n I_q(U_{1,i};Y_{1,i}) \,,
\label{eq:convBcompound5}
\end{align}
where the last equality holds since $U_{0,i}\Cbar U_{1,i}\Cbar Y_{1,i}$ form a Markov chain. As for (\ref{eq:convBcompound3}), we have that
\begin{align}
I_q(M_1;Y_1^n|M_0)+I_q(M_0,Y_2^n) 
=&\sum_{i=1}^n I_q(M_1;Y_{1,i}|M_0,Y_1^{i-1})+\sum_{i=1}^n I_q(M_0;Y_{2,i}|Y_{2,i+1}^n) \nonumber \\
\leq& \sum_{i=1}^n I_q(M_1,Y_{2,i+1}^n;Y_{1,i}|M_0,Y_1^{i-1})+\sum_{i=1}^n I_q(M_0,Y_{2,i+1}^n;Y_{2,i}) \nonumber\\
=& \sum_{i=1}^n I_q(M_1;Y_{1,i}|M_0,Y_1^{i-1},Y_{2,i+1}^n)+\sum_{i=1}^n I_q(Y_{2,i+1}^n;Y_{1,i}|M_0,Y_1^{i-1}) \nonumber\\
&+\sum_{i=1}^n I_q(M_0,Y_1^{i-1},Y_{2,i+1}^n;Y_{2,i})-\sum_{i=1}^n I_q(Y_1^{i-1};Y_{2,i}|M_0,Y_{2,i+1}^n) 
\,.
\label{eq:convBcompound6}
\end{align}
Then, the second and fourth sums cancel out, by the Csisz{\'a}r sum identity \cite[Section 2.3]{ElGamalKim:11b}.   Hence,
\begin{align}
I_q(M_1;Y_1^n|M_0)+I_q(M_0;Y_2^n)  
\leq& \sum_{i=1}^n I_q(M_1;Y_{1,i}|M_0,Y_1^{i-1},Y_{2,i+1}^n)
+\sum_{i=1}^n I_q(M_0,Y_1^{i-1},Y_{2,i+1}^n;Y_{2,i}) \nonumber\\
\leq& \sum_{i=1}^n I_q(U_{1,i};Y_{1,i}|U_{0,i})+\sum_{i=1}^n I_q(U_{0,i};Y_{2,i})
\,.
\label{eq:convBcompound7}
\end{align}
Thus, by (\ref{eq:convBcompound1})--(\ref{eq:convBcompound3}) and (\ref{eq:convBcompound5})--(\ref{eq:convBcompound7}), we have that 
\begin{align}
R_0\leq& \frac{1}{n}\sum_{i=1}^n I_q(U_{0,i};Y_{2,i})+\eps_n \,, \label{eq:convBcompound8}
\\
R_0+R_1\leq& \frac{1}{n}\sum_{i=1}^n I_q(U_{1,i};Y_{1,i})+\eps_n \,, \label{eq:convBcompound9}
\\
R_0+R_1\leq& \frac{1}{n}\sum_{i=1}^n I_q(U_{1,i};Y_{1,i} |U_{0,i})+\sum_{i=1}^n I_q(U_{0,i};Y_{2,i})+ \eps_n \,. \label{eq:convBcompound10}
\end{align}
Introducing a time-sharing random variable $K$, uniformly distributed over $[1:n]$ and independent of $(S^n,U_0^n,U_1^n)$, we have that 
\begin{align}
R_0 \leq& I_q(U_{0,K};Y_{2,K}|K)+\eps_n \,,\label{eq:convBcompound11}\\
R_0+R_1\leq& I_q(U_{1,K};Y_{1,K}|K) +\eps_n \,,\label{eq:convBcompound12}\\
R_0+R_1\leq& I_q(U_{1,K};Y_{1,K}|U_{0,K},K) +I_q(U_{0,K};Y_{2,K}|K)+\eps_n \,. \label{eq:convBcompound13}
\end{align}
Define $U_0\triangleq (U_{0,K},K)$ and $U_1\triangleq (U_{1,K},K)$. Hence, 
$P_{Y_{1,K},Y_{2,K}|U_0,U_1}=P_{Y_{1},Y_{2}|U_0,U_1}$. Then, 
by (\ref{eq:BcompConvEquiv2}) and (\ref{eq:convBcompound11})--(\ref{eq:convBcompound13}), it follows that 
$(R_0,R_1)\in \BICcompound$. 
\end{proof}

\section*{Part 2}
We show that when  the set of state distributions $\Qset$ is convex,  and Condition $\sCondQ$ holds, 
the capacity region of the compound broadcast channel $\Bcompound$ with causal SI is given by $\BCcompound=\BrCcompound=\BIRcompound=\BICcompound$ 
(and this holds regardless of whether the interior of the capacity region is empty or not). 

Due to part 1, we have that
\begin{align}
\label{eq:Bcompound2up}
 \BrCcompound \subseteq \BICcompound \,.
\end{align} 
By Lemma~\ref{lemm:BcompoundLowerB},
\begin{align}
\BCcompound \supseteq \BIRcompound \,.
\end{align}
Thus, 
\begin{align}
\label{eq:BcompoundTightineq}
 \BIRcompound\subseteq  \BCcompound \subseteq \BrCcompound \subseteq \BICcompound \,. 
\end{align}

To conclude the proof, we show that Condition $\sCondQ$ implies that $\BIRcompound\supseteq\BICcompound$, hence the inner and outer bounds coincide. By Definition~\ref{def:Bcompoundachieve}, if a function $\encs(u_0,u_1,s)$ and a set 
$\Dset$ achieve $\BIRcompound$ and $\BICcompound$, then
\begin{subequations}
\label{eq:BcompoundachieveEq} 
\begin{align}  
\label{eq:BIRcompoundachieveEq} 
\BIRcompound =\bigcup_{p(u_0,u_1)\in\Dset}\,  
\left\{
\begin{array}{lrl}
(R_0,R_1) \,:\; & R_0 &\leq  \min_{q\in\Qset} I_q(U_0;Y_2) \,, \\
								& R_1 &\leq  \min_{q\in\Qset} I_q(U_1;Y_1|U_0) \,,\\
								& R_0+R_1&\leq \min_{q\in\Qset} I_q(U_0,U_1;Y_1)
\end{array}
\right\} \,,
\intertext{and}
\label{eq:BICcompoundachieveEq} 
\BICcompound = \bigcap_{q(s)\in\Qset}\, \bigcup_{p(u_0,u_1)\in\Dset} 
\left\{
\begin{array}{lrl}
(R_0,R_1) \,:\; & R_0 &\leq   I_q(U_0;Y_2) \,, \\
								& R_1 &\leq   I_q(U_1;Y_1|U_0)  \,,\\
								& R_0+R_1&\leq  I_q(U_0,U_1;Y_1)
\end{array}
\right\} \,.
\end{align}
\end{subequations}
Hence, when Condition $\sCondQ$ holds, we have  by Definition~\ref{def:sCondQ} 
 that for some  $\encs(u_0,u_1,s)$, $\Dset\subseteq\pSpace(\Uset_0\times\Uset_1)$, and $q^*\in\Qset$,
\begin{align}
\BIRcompound=&  \bigcup_{p(u_0,u_1)\in\Dset}
\left\{
\begin{array}{lrl}
(R_0,R_1) \,:\; & R_0 &\leq   I_{q^*}(U_0;Y_2) \,, \\
								& R_1 &\leq   I_{q^*}(U_1;Y_1|U_0)  \,,\\
								& R_0+R_1&\leq  I_{q^*}(U_0,U_1;Y_1)
\end{array}
\right\} \nonumber\\
\supseteq& \BICcompound \,,
\end{align}
where the last line follows from (\ref{eq:BICcompoundachieveEq}).
%
%
\qed

\section{Proof of Theorem~\ref{theo:BCrp}}
\label{app:BCrp}
At first,  ignore the cardinality bounds in (\ref{eq:BICrpAlph}). Then, it immediately follows from Theorem~\ref{theo:BcvC} that $\BCrp=\BICrp$, by taking the set $\Qset$ that consists of a single state distribution $q(s)$. 

To prove the bounds on the alphabet sizes of the strategy variables $U_0$ and $U_1$, we apply the standard Carath{\'e}odory techniques (see \eg \cite[Lemma 15.4]{CsiszarKorner:82b}).
Let
\begin{align}
\label{eq:BCrpL0}
L_0\triangleq (|\Xset|-1)|\Sset|+3 \leq |\Xset| |\Sset|+2 \,,
\end{align}
where the inequality holds since $|\Sset|\geq 1$. 
 Without loss of generality, assume that $\Xset=[1:|\Xset|]$ and $\Sset=[1:|\Sset|]$. 
Then, define the following $L_0$ functionals,
\begin{align}
 &\varphi_{ij}(P_{U_1,X|S})=\sum\limits_{u_1\in\Uset_1} P_{U_1,X|S}(u_1,i|j)= P_{X|S}(i|j) \,,\; i=1,\ldots |\Xset|-1, 
j=1,\ldots, |\Sset| \,, \\
&\psi_{1}(P_{U_1,X|S})= 
-\sum\limits_{s,u_1,x,y_1} q(s) P_{U_1,X|S}(u_1,x|s) W_{Y_1|X,S}(y_1|x,s) 
 \log \left[ 
\sum\limits_{s',u_1',x'} q(s') P_{U_1,X|S}(u_1',x'|s') W_{Y_1|X,S}(y_1|x',s')      \right] \,,\\
 &\psi_{2}(P_{U_1,X|S})= 
-\sum\limits_{s,u_1,x,y_2} q(s) P_{U_1,X|S}(u_1,x|s) W_{Y_2|X,S}(y_2|x,s) 
 \log \left[ 
\sum\limits_{s',u_1',x'} q(s') P_{U_1,X|S}(u_1',x'|s') W_{Y_2|X,S}(y_2|x',s')      \right] \,,\\
 &\psi_{3}(P_{U_1,X|S})=
-\sum\limits_{u_1,x,s} q(s)P_{U_1,X|S}(u_1,x|s) \log\left[ \sum_{x',s'} q(s')P_{U_1,X|S}(u_1,x'|s')		\right]
\nonumber\\ 
&-\sum\limits_{u_1,x,s} q(s)P_{U_1,X|S}(u_1,x|s)W_{Y_1|X,S}(y_1|x,s)
 \log\left[ \frac{ \sum_{x',s'} q(s')P_{U_1,X|S}(u_1,x'|s')W_{Y_1|X,S}(y_1|x',s') }{\sum_{u_1'',x'',s''} q(s'')P_{U_1,X|S}(u_1'',x''|s'')W_{Y_1|X,S}(y_1|x'',s'')}
		\right] \,.
\end{align}
Then, observe that 
\begin{align}
&\sum_{u_0\in\Uset_0} p(u_0) \varphi_{i,j}(P_{U_1,X|S,U_0}(\cdot,\cdot|\cdot,u_0))= P_{X|S}(i|j) \,,\\
&\sum_{u_0\in\Uset_0} p(u_0) \psi_{1}(P_{U_1,X|S,U_0}(\cdot,\cdot|\cdot,u_0))= H(Y_1|U_0) \,,\\
&\sum_{u_0\in\Uset_0} p(u_0) \psi_{1}(P_{U_1,X|S,U_0}(\cdot,\cdot|\cdot,u_0))= H(Y_2|U_0) \,,\\
&\sum_{u_0\in\Uset_0} p(u_0) \psi_{1}(P_{U_1,X|S,U_0}(\cdot,\cdot|\cdot,u_0))= I(U_1;Y_1|U_0) \,.
\end{align}
By \cite[Lemma 15.4]{CsiszarKorner:82b}, the alphabet size of $U_0$ can then be restricted to $|\Uset_0|\leq L_0$, while preserving $P_{X,S,Y_1,Y_2}$; $I(U_0;Y_2)=H(Y_2)-H(Y_2|U_0)$; $I(U_0;Y_1|U_0)$; and
$I(U_0,U_1;Y_1)=I(U_0;Y_1|U_0)+H(Y_1)-H(Y_1|U_0)$. 

Fixing the alphabet of $U_0$, we now apply similar arguments to the cardinality of $\Uset_1$.
Then, less than 
$|\Xset| |\Sset| L_0 -1$
 functionals are required for the joint distribution $P_{U_0,X|S}$,
 and an additional functional to preserve $H(Y_1|U_1,U_0)$. Hence, by \cite[Lemma 15.4]{CsiszarKorner:82b}, the alphabet size of $U_0$ can then be restricted to $|\Uset_1|\leq |\Xset| |\Sset| L_0\leq |\Xset| |\Sset| (|\Xset| |\Sset|+2)$ (see (\ref{eq:BCrpL0})).
\qed

\section{Proof of Theorem~\ref{theo:Bmain}}
\label{app:Bmain}

\subsection{Part 1}
First, we explain the general idea. 
We devise  a causal version of Ahlswede's Robustification Technique 
(RT)  \cite{Ahlswede:86p,WinshtokSteinberg:06c}. Namely, we use codes for the compound  broadcast channel to construct a random code for the AVBC using randomized permutations. However, in our case, the causal nature of the problem imposes a difficulty, and the application of the RT is not straightforward.

In \cite{Ahlswede:86p,WinshtokSteinberg:06c}, the state information is noncausal and a random code is defined via permutations of the codeword symbols. This cannot be done here, because the SI is provided to the encoder in a causal manner. 
We resolve this difficulty using Shannon strategy codes for the compound  broadcast channel to construct a random code for the AVBC, applying permutations to the \emph{strategy sequence} $(u_1^n,u_0^n)$, which is an integral part of the Shannon strategy code, and is independent of the channel state. The details are given below.

\subsubsection{Inner Bound}
We show that the region defined in (\ref{eq:BIRcompoundP}) can be achieved by random codes over the 
AVBC $\avbc$ with causal SI, \ie $\BCavc \supseteq \BIRavc$.
We start with Ahlswede's RT \cite{Ahlswede:86p}, stated below. Let $h:\Sset^n\rightarrow [0,1]$ be a given function. If, for some fixed $\alpha_n\in(0,1)$, and for all 
$ \qn(s^n)=\prod_{i=1}^n q(s_i)$, with 
$q\in\pSpace(\Sset)$, 
\begin{align}
\label{eq:BRTcondC}
\sum_{s^n\in\Sset^n} \qn(s^n)h(s^n)\leq \alpha_n \,,
\end{align}
then,
\begin{align}
\frac{1}{n!} \sum_{\pi\in\Pi_n} h(\pi s^n)\leq \beta_n \,,\quad\text{for all $s^n\in\Sset^n$} \,,
\end{align}
where $\Pi_n$ is the set of all $n$-tuple permutations $\pi:\Sset^n\rightarrow\Sset^n$, and 
$\beta_n=(n+1)^{|\Sset|}\cdot\alpha_n$. 

According to Lemma~\ref{lemm:BcompoundLowerB}, 
 for every $(R_0,R_1)\in\BIRavc$, there exists a  $(2^{nR_0},$ $2^{nR_1},$ $n,$ $e^{-2\theta n})$ Shannon strategy code for the compound broadcast channel $\BcompoundP$ with causal SI, for some $\theta>0$ and sufficiently large $n$. 
Given such a Shannon strategy code $\code=$ $(u_0^n(m_0),$ $u_1^n(m_0,m_1),$  $\encs(u_0,u_1,s),$ $\dec_1(y_1^n),$ $\dec_2(y_2^n))$,
 we have that (\ref{eq:BRTcondC}) is satisfied with  $h(s^n)=\cerr(\code)$  and $\alpha_n=e^{-2\theta n}$.  
As a result, Ahlswede's RT tells us that
\begin{align}
\label{eq:BdetErrC}
\frac{1}{n!} \sum_{\pi\in\Pi_n} P_{e|\pi s^n}^{(n)}(\code)\leq (n+1)^{|\Sset|}e^{-2\theta n} 
\leq e^{-\theta n}  \,,\quad\text{for all $s^n\in\Sset^n$} \,,
\end{align} 
for a sufficiently large $n$, such that $(n+1)^{|\Sset|}\leq e^{\theta n}$.  

On the other hand, for every $\pi\in\Pi_n$,  
\begin{align}
P_{e|\pi s^n}^{(n)}(\code)			
 &\stackrel{(a)}{=}
\frac{1}{2^{ n(R_0+R_1) }}\sum_{m_0,m_1}
\sum_{(\pi y_1^n,\pi y_2^n)\notin\Dset(m_0,m_1)} 
\nBC(\pi y_1^n,\pi y_2^n|\encs^n(u_0^n(m_0),u_1^n(m_0,m_1),\pi s^n),\pi s^n) 	\nonumber\\
 &\stackrel{(b)}{=}
\frac{1}{2^{ n(R_0+R_1) }}\sum_{m_0,m_1}
\sum_{(\pi y_1^n,\pi y_2^n)\notin\Dset(m_0,m_1)} 
\nBC( y_1^n, y_2^n|\pi^{-1} \encs^n(u_0^n(m_0),u_1^n(m_0,m_1),\pi s^n), s^n) \,,\nonumber\\
 &\stackrel{(c)}{=}\frac{1}{2^{ n(R_0+R_1) }}\sum_{m_0,m_1}
\sum_{
(\pi y_1^n,\pi y_2^n)\notin\Dset(m_0,m_1)}
\nBC( y_1^n, y_2^n| \encs^n(\pi^{-1} u_0^n(m_0), \pi^{-1} u_1^n(m_0,m_1), s^n), s^n)
\label{eq:Bcerrpi}
\end{align}
where $(a)$ is obtained by  plugging $\pi s^n$ and $x^n=\encs^n(\cdot,\cdot,\cdot)$ in (\ref{eq:Bcerr}) 
 and then changing the order of summation over $(y_1^n,y_2^n)$; $(b)$ holds because the broadcast channel is memoryless; and $(c)$ follows from that fact that  for a Shannon strategy code,  $x_i=\encs(u_{0,i},u_{1,i},s_i)$, $i\in[1:n]$, by Definition~\ref{def:BStratCode}. 
The last expression suggests the use of permutations applied to the encoding \emph{strategy sequence} and the channel output sequences.

Then, consider the $(2^{nR_0},2^{nR_1},n)$ random code $\code^\Pi$, specified by 
\begin{subequations}
\label{eq:BCpi}
\begin{align}
f_\pi^n(m_0,m_1,s^n)&= \encs^n(\pi^{-1} u_1^n(m_0,m_1),\pi^{-1} u_0^n(m_0),s^n) \,,\\
\intertext{and}
g_{1,\pi}(y_1^n)&=\dec_1(\pi y_1^n)
\,,\quad g_{2,\pi}(y_2^n)=\dec(\pi y_2^n) \,,
\end{align}
\end{subequations}
for $\pi\in\Pi_n$,
with a uniform distribution $\mu(\pi)=\frac{1}{|\Pi_n|}=\frac{1}{n!}$. 
Such permutations can be implemented without knowing $s^n$, hence this coding scheme does not violate the causality requirement. 

 From (\ref{eq:Bcerrpi}), 
 we see that 
\begin{align} 
\cerr(\code^\Pi)=\sum_{\pi\in\Pi_n} \mu(\pi) P_{e|\pi s^n}^{(n)}(\code) \,,
\end{align}
for all $s^n\in\Sset^n$, and therefore, together with (\ref{eq:BdetErrC}), we have that the probability of error of the random code $\code^\Pi$ is bounded by 
\begin{align} 
\err(\qn,\code^{\Pi})\leq e^{-\theta n} \,,
\end{align} 
for every $\qn(s^n)\in\pSpace(\Sset^n)$. That is, $\code^\Pi$ is a $(2^{nR_0},2^{nR_1},n,e^{-\theta n})$ random 
 code for the AVBC $\avbc$ with causal SI at the encoder. 
This completes the proof of the inner bound. 
\qed 

\subsubsection{Outer Bound}
We show that the capacity region of the AVBC $\avbc$ with causal SI is bouned by 
 $\BrCav\subseteq \BrICav$ (see (\ref{eq:BIRcompoundP})). 
%
The random code capacity region of the AVBC is included within the random code capacity region of the compound  broadcast channel, namely
\begin{align}
\label{eq:Outercomp2}
\BrCav \subseteq \BrCcompoundP \,.
\end{align}
By Theorem~\ref{theo:BcvC} we have that $\BrCcompound\subseteq\BICcompound$. Thus,   
with $\Qset=\pSpace(\Sset)$,
\begin{align}
\label{eq:Outercomp11}
\BrCcompoundP\subseteq\BrICav \,.
\end{align}
It follows from (\ref{eq:Outercomp2}) and (\ref{eq:Outercomp11}) that $\BrCav\subseteq\BrICav$. Since the random code capacity region always includes  the deterministic code capacity region, we have that
$
\BCavc \subseteq \BrICav 
$ as well.
\qed 

\subsection*{Part 2}
The second equality, $\BIRavc=\BrICav$, follows from  part 2 of Theorem~\ref{theo:BcvC}, taking $\Qset=\pSpace(\Sset)$. 
By part 1, $\BIRavc\subseteq \BrCav\subseteq\BrICav$, hence the proof follows. \qed

\section{Proof of Lemma~\ref{lemm:BcorrSizeC}}
\label{app:BET}

The proof follows the lines of \cite[Section~4]{Ahlswede:78p}. 
 Let $\dK>0$ be an integer, chosen later, and define the random variables
\begin{align}
\label{eq:BLi}
L_1,L_2,\ldots,L_{\dK} \;\,\text{i.i.d. $\sim\mu(\ell)$} \;.
\end{align}
Fix $s^n$, and define the random variables
\begin{align}
\Omega_j(s^n)= \cerr(\code_{L_j}) \;,\quad j\in [1:\dK] \;,
\end{align}
which is the conditional probability of error of the code $\code_{L_j}$ given the state sequence $s^n$. 

Since $\code^\Gamma$ is a $(2^{nR_1},2^{nR_2},n,\eps_n)$ code, we have that 
$ \sum_\gamma\mu(\gamma)\sum_{s^n} \qn(s^n) \cerr(\code_\gamma)\leq \eps_n$, for all $\qn(s^n)$. In particular, for a kernel, we have that  
\begin{align}
\label{eq:BPsiIneqC}
\E \Omega_j(s^n)=\sum_{\gamma\in\Gamma} \mu(\gamma)\cdot \cerr(\code_\gamma) \leq \eps_n \;,
\end{align}
for all $j\in[1:k]$.

Now take $n$ to be large enough so that $\eps_n<\alpha$. 
Keeping $s^n$ fixed,  we have that the random variables $\Omega_j(s^n)$ are i.i.d., due to (\ref{eq:BLi}).
{ Next the technique known as Bernstein's trick \cite{Ahlswede:78p} is applied. }
\begin{align}
\prob{\sum_{j=1}^{\dK} \Omega_j(s^n)\geq \dK\alpha} \stackrel{(a)}{\leq}&
\E\left\{ \exp\left[ \beta \left(
\sum_{j=1}^\dK \Omega_j(s^n)- k\alpha
\right)
\right]
\right\} \\
=& e^{-\beta \dK \alpha}\cdot
\E\left\{  \prod_{j=1}^\dK e^{\beta  \Omega_j(s^n)} \right\} \\
\stackrel{(b)}{=}& e^{-\beta \dK \alpha}\cdot
\prod_{j=1}^\dK \E\left\{   e^{\beta  \Omega_j(s^n)} \right\} \\
\stackrel{(c)}{\leq}& e^{-\beta \dK \alpha}\cdot
\prod_{j=1}^\dK \E\left\{   1+e^\beta \cdot\Omega_j(s^n) \right\} \\
\stackrel{(d)}{\leq}& e^{-\beta \dK \alpha}\cdot
 \left(   1+e^\beta \eps_n \right)^\dK 
\end{align}
where $(a)$ is an application of Chernoff's inequality; $(b)$ follows from the fact that $\Omega_j(s^n)$ are independent; $(c)$ holds since $e^{\beta x}\leq1+e^\beta x$, for $\beta>0$ and $0\leq x\leq 1$; $(d)$ follows from (\ref{eq:BPsiIneqC}). We take $n$ to be large enough for $1+e^\beta \eps_n\leq e^\alpha$ to hold. Thus, choosing $\beta=2$, we have that 
\begin{align}
\prob{\frac{1}{\dK}\sum_{j=1}^\dK \Omega_j(s^n)\geq \alpha} \leq&
e^{- \alpha k } \;,
\end{align}
for all $s^n\in\Sset^n$. Now, by the union of events bound, we have that
\begin{align}
\prob{
\max_{s^n}  \frac{1}{\dK} \sum_{j=1}^\dK \Omega_j(s^n) \,\geq\, \alpha
 }=&\prob{
\exists s^n :  \frac{1}{\dK} \sum_{j=1}^\dK \Omega_j(s^n) \,\geq\, \alpha
 }\\
\leq&\sum_{s^n\in\Sset^n} \prob{  \frac{1}{\dK} \sum_{j=1}^\dK \Omega_j(s^n) \,\geq\, \alpha
 }\\
\leq& |\Sset|^n \cdot e^{-\alpha \dK} \;.
\label{eq:BeBound}
\end{align}
Since $|\Sset|^n$ grows only exponentially in $n$, choosing $k=n^2$ results in a super exponential decay. 

Consider the code $\code^{\Gamma^*}=(\mu^*,\Gamma^*=[1:k],\{\code_{L_j}\}_{j=1}^k)$ formed by a random collection of codes, with $\mu^*(j)=\frac{1}{k}$.  It follows 
 that  the conditional  probability of error given  $s^n$, which is given by  
\begin{align} 
\cerr(\code^{\Gamma^*})=
\frac{1}{k} \sum_{j=1}^{k} \cerr(\code_{L_j}) \;,
\end{align}
 exceeds $\alpha$ with a super exponentially small probability $\sim e^{-\alpha n^2}$, for all $s^n\in\Sset^n$.
 Thus, there exists a random code $\code^{\Gamma^*}=(\mu^*,\Gamma^*,\{\code_{\gamma_j}\}_{j=1}^k)$ 
 for the AVBC $\avbc$, such that 
\begin{align}
\err(\qn, \code^{\Gamma^*})=\sum_{s^n\in\Sset^n} \qn(s^n) \cerr(\code^{\Gamma^*})\leq \alpha \;,\quad\text{for all $\qn(s^n)
\in\pSpace(\Sset^n)$}\;.
\end{align}

\qed

\section{Proof of Theorem~\ref{theo:BcorrTOdetC}}
\label{app:BcorrTOdetC}
\begin{proof}[Achievability proof]
 To show achievability, 
we follow the lines of \cite{Ahlswede:78p}, with the required adjustments.
  We use the random code constructed in the proof of  Theorem~\ref{theo:Bmain} 
	to construct a deterministic code.

Let $(R_0,R_1)\in\BrCav$, and consider the case where $\interior{\BCavc}\neq \emptyset$. Namely,
\begin{align}
\label{eq:Brpos}
\opC(\avc_1)>0 \,,\;\text{and}\; \opC(\avc_2)>0 \,,
\end{align}
where $\avc_1=\{\sbc\}$ and $\avc_2=\{\wbc\}$ denote the marginal AVCs with causal SI of 
user 1 and user 2, 
respectively.  
By Lemma~\ref{lemm:BcorrSizeC},  for every $\eps_1>0$ and sufficiently large $n$, 
 there exists a $(2^{nR_0},2^{nR_1},n,\eps_1)$ random  code  
$
\code^\Gamma=\big(\mu(\gamma)=\frac{1}{k},\Gamma=[1:k],\{\code_\gamma \}_{\gamma\in \Gamma}\big) 
$, 
where
$\code_\gamma=(\encn_\gamma,\dec_{1,\gamma},\dec_{2,\gamma})$, 
for $\gamma\in\Gamma$, 
 and 
$
k=|\Gamma|\leq n^2 
$. 
Following (\ref{eq:Brpos}),  we have that for every $\eps_2>0$ and sufficiently large $\nu$, the code index $\gamma\in [1:k]$ can be sent over $\avbc$ using a $(2^{\nu\bR_0},2^{\nu\bR_1},\nu,\eps_2)$ deterministic code 
$ 
\code_{\text{i}}=(\tfnu,\gnu_1,\gnu_0)  
$, where $\bR_0>0$, $\bR_1>0$.
Since $k$ is at most polynomial, 
 the encoder can reliably convey $\gamma$ to the receiver with a negligible blocklength, \ie
$ 
\nu=o(n) 
$. 

Now, consider a code 
 formed by the concatenation of $\code_{\text{i}}$ as a prefix to a corresponding code in the code collection $\{\code_\gamma\}_{\gamma\in\Gamma}$. 
That is, the encoder sends both the index $\gamma$ and the message pair $(m_0,m_1)$ to the receivers, such that 
 the index $\gamma$ is transmitted first by $\tfnu(\gamma,s^\nu)$, and then the message pair $(m_0,m_1)$ is transmitted by the codeword $x^n=\enc_\gamma^n($ $m_0,m_1,$ $s_{\nu+1},\ldots,s_{\nu+n})$.
Subsequently, decoding is performed in two stages as well; decoder 1 estimates the index at first, with  
$\hgamma_1=$ $\gnu_1(y_{1,1},\ldots,$ $y_{1,\nu})$, and the message pair $(m_0,m_1)$ is then estimated by  
$(\widehat{m}_0,\widehat{m}_1)=$ $g_{1,\hgamma_1}(y_{1,\nu+1},$ $\ldots,y_{1,\nu+n})$.  
Similarly, decoder 2 estimates the index  with  
$\hgamma_2=$ $\gnu_0(y_{2,1},$ $\ldots,y_{2,\nu})$, and the message $m_0$ is then estimated by  
$\widetilde{m}_2=$ $g_{2,\hgamma_2}(y_{2,\nu+1},\ldots,$ $y_{2,\nu+n})$.

 By the union of events bound, the probability of error 
 is then bounded by $\eps=\eps_1+\eps_2$, 
for every joint distribution 
in $\pSpace^{\nu+n}(\Sset^{\nu+n})$. 
That is, the concatenated code 
 is a $(2^{(\nu+n)\tR_{1,n}},2^{(\nu+n)\tR_{2,n}},\nu+n,\eps)$ code over the AVBC $\avbc$ with causal SI, where $\nu=o(n)$. 
Hence, 
  the blocklength is $n+o(n)$, and 
the rates   $\tR_{0,n}=\frac{n}{\nu+n}\cdot R_0$ and $\tR_{1,n}=\frac{n}{\nu+n}\cdot R_1$ approach $R_0$ and 
$R_1$, respectively, as $n\rightarrow \infty$. 
\end{proof}

\begin{proof}[Converse proof]
In general, the deterministic code capacity region is included within the random code capacity region. Namely,
$\BCavc\subseteq\BrCav$. 
\end{proof} 

\section{Proof of Corollary~\ref{coro:BmainDbound}}
\label{app:BmainDbound}
First, consider the inner and outer bounds in (\ref{eq:BmainInner}) and (\ref{eq:BmainOuter}).
The bounds are obtained as a  direct consequence of part 1 of Theorem~\ref{theo:Bmain} and  Theorem~\ref{theo:BcorrTOdetC}. Note that the outer bound (\ref{eq:BmainOuter}) holds regardless of any condition, since the deterministic code capacity region is always included within the random code capacity region, \ie $\BCavc \subseteq \BrCav\subseteq \BrICav$.

Now, suppose that the marginals $V_{Y_1|U,S}^{\encs}$ and $V_{Y_2|U_0,S}^{\encs'}$ are non-symmetrizable for some $\encs:
\Uset\times\Sset\rightarrow\Xset$ and $\encs':
\Uset_0\times\Sset\rightarrow\Xset$, and Condition $\sCond$ holds. 
 Then, based on \cite{CsiszarNarayan:88p,CsiszarKorner:82b}, 
both marginal (single-user) AVCs have positive capacity, \ie $\opC(\avc_1)>0$ and $\opC(\avc_2)>0$. Namely,
 $\interior{\BCavc}\neq\emptyset$. 
%
 Hence, by Theorem~\ref{theo:BcorrTOdetC}, the deterministic code capacity region coincides with the random code capacity region, \ie $\BCavc=\BrCav$. Then, the proof follows from part 2 of Theorem~\ref{theo:Bmain}. \qed 


\section{Analysis of Example~\ref{example:AVBSBC}}
\label{app:AVBSBC}
We begin with the case of an arbitrarily varying BSBC $\avbc_{D,0}$ without SI. 
We claim that the single user marginal AVC $\avc_{1,0}$  without SI, corresponding to the stronger user, has zero capacity.
Denote $q\triangleq q(1)=1-q(0)$. 
Then, observe that the additive noise is distributed according to $Z_S\sim\text{Bernoulli}(\eps_q)$, with
$
\eta_q\triangleq (1-q)\cdot \theta_0+q\cdot \theta_1 
$, 
for $0\leq q\leq 1$. 
Based on \cite{BBT:60p}, 
$\opC(\avc_{1,0})\leq\opC^{\rstarC}\hspace{-0.1cm}(\avc_{1,0})= \min_{0\leq q\leq 1} [1-h(\eta_q)]$.
Since $\theta_0< \frac{1}{2}\leq \theta_1$, there exists $0\leq q\leq 1$
such that $\eta_q=\frac{1}{2}$, thus $\opC(\avc_{1,0})=0$.
%
The capacity region of the AVDBC $\avbc_{D,0}$ without SI is then given by  
$\sBCavcig=\{(0,0)\}$.

Now, consider the arbitrarily varying BSBC $\savbc$ with causal SI. 
 By Theorem~\ref{theo:Bmain1}, 
 the random code capacity region is bounded by 
$
\sBIRavc \subseteq \sBrCav\subseteq \sBrICav 
$. 
We show that the bounds coincide, and are thus tight. 
Let $\Brp_D$ denote the random parameter DBC $\bc$ with causal SI, governed by  an i.i.d. state sequence, distributed according to $S\sim\text{Bernoulli}(q)$.
By \cite{Steinberg:05p}, the corresponding capacity region 
 is given by
\begin{subequations}
\label{eq:Bex1ICrpT}
\begin{align}
\label{eq:Bex1ICrpa}
&\inC(\savbc^{q})=
\bigcup_{0\leq \beta\leq 1}
\left\{
\begin{array}{lll}
(R_1,R_2) \,:\; & R_2 &\leq   1-h(\alpha*\beta*\delta_q) \,, \\
								& R_1 &\leq   h(\beta*\delta_q)-h(\delta_q)
\end{array}
\right\}
 \,,
\end{align}
where
 \begin{align}
\label{eq:ex1deltaq}
\delta_q\triangleq (1-q)\cdot \theta_0+q\cdot (1-\theta_1) \,,
\end{align}
\end{subequations}
 for $0\leq q\leq 1$. For every given $0\leq q'\leq 1$, we have that 
$
\sBrICav=\bigcap_{0\leq q\leq 1} \inC(\savbc^{q}) \subseteq 
\inC(\savbc^{q'}) 
$. 
 Thus, taking $q'=1$, we have that
\begin{align}
\label{eq:ex1out}
\sBrICav
 \subseteq 
\bigcup_{0\leq \beta\leq \frac{1}{2}}
\left\{
\begin{array}{lll}
(R_1,R_2) \,:\; & R_2 &\leq   1-h(\alpha*\beta*\theta_1) \,, \\
								& R_1 &\leq   h(\beta*\theta_1)-h(\theta_1)
\end{array}
\right\}\,,
\end{align}
where we have used the identity 
$h(\alpha*(1-\delta))=h(\alpha*\delta)$. 

Now, to show that the region above is achievable, we examine the inner bound,
\begin{align}
\sBIRavc=
\bigcup_{p(u_1,u_2),\encs(u_1,u_2,s)}\, 
\left\{
\begin{array}{lll}
(R_1,R_2) \,:\; & R_2 &\leq \min_{0\leq q\leq 1}   I_q(U_2;Y_2) \,, \\
								& R_1 &\leq \min_{0\leq q\leq 1}  I_q(U_1;Y_1|U_2)  
\end{array}
\right\}\,.
\end{align}
Consider the following choice of $p(u_1,u_2)$ and $\encs(u_1,u_2,s)$. Let $U_1$ and $U_2$ be independent random variables, 
\begin{align}
\label{eq:BSBCdistAchieve}
U_1\sim\text{Bernoulli}(\beta) \,,\;\text{and}\;\, U_2\sim\text{Bernoulli}\left(\frac{1}{2} \right)\,,
\end{align}
for $0\leq \beta\leq\frac{1}{2}$, 
 and let 
\begin{align}
\label{eq:BSBCxiAchieve}
\encs(u_1,u_2,s)=u_1+u_2+s \mod 2\,.
\end{align}
Then,
\begin{align}
 &H_q(Y_1|U_1,U_2)=H_q(S+Z_S)=h(\delta_q) \,,																\nonumber\\ 
 &H_q(Y_1|U_2)=H_q(U_1+S+Z_S)=h(\beta*\delta_q) \,,													\nonumber\\
 &H_q(Y_2|U_2)=H_q(U_1+S+Z_S+V)=h(\alpha*\beta*\delta_q) \,,								\nonumber\\ 
 &H_q(Y_2)=1 \,, 
\intertext{
where addition is modulo $2$, and $\delta_q$ is given by (\ref{eq:ex1deltaq}). 
Thus,
}
&I_q(U_2;Y_2)=1-h(\alpha*\beta*\delta_q) \,, \nonumber\\
&I_q(U_1;Y_1|U_2)=h(\beta*\delta_q)-h(\delta_q) \,,
\end{align}
hence 
\begin{align}
\label{ex1:innerR}
\sBIRavc \supseteq
\bigcup_{0\leq \beta\leq \frac{1}{2}}\, 
\left\{
\begin{array}{lll}
(R_1,R_2) \,:\; & R_2 &\leq \min_{0\leq q\leq 1}  1-h(\alpha*\beta*\delta_q) \,, \\
								& R_1 &\leq \min_{0\leq q\leq 1}  h(\beta*\delta_q)-h(\delta_q)  
\end{array}
\right\}\,.
\end{align}
Note that $\theta_0\leq \delta_q\leq 1-\theta_1 \leq\frac{1}{2}$. For $0\leq\delta\leq\frac{1}{2}$, the functions 
$g_1(\delta)=1-h(\alpha*\beta*\delta)$ and $g_2(\delta)= h(\beta*\delta)-h(\delta)$ are monotonic decreasing functions of $\delta$, hence the minima in (\ref{ex1:innerR}) are both achieved with $q=1$.
It follows that 
	\begin{align}
	\label{eq:ex1BrCav}
\sBrCav=\sBIRavc=\sBrICav=
\bigcup_{0\leq \beta\leq 1}\, 
\left\{
\begin{array}{lll}
(R_1,R_2) \,:\; & R_2 &\leq   1-h(\alpha*\beta*\theta_1) \,, \\
								& R_1 &\leq h(\beta*\theta_1)-h(\theta_1)  
\end{array}
\right\}\,.
\end{align}

It can also be verified that Condition $\sCond_D$ holds (see Definition~\ref{def:sCond1}), in agreement with part 2 of Theorem~\ref{theo:Bmain1}. 
First, we specify a function $\encs(u_1,u_2,s)$ and a distributions set 
$\Dset^{\rstarC}$ that achieve $\sBIRavc$ and $\sBrICav$ (see Definition~\ref{eq:Bachieve1}).
 Let $\encs(u_1,u_2,s)$ be as in (\ref{eq:BSBCxiAchieve}), and let $\Dset^{\rstarC}$ be the set of distributions $p(u_1,u_2)$ such that $U_1$ and $U_2$ are independent random variables, distributed according to  (\ref{eq:BSBCdistAchieve}). 
By the derivation above,  the requirement (\ref{eq:BIRachieve1}) is satisfied. Now, by the derivation in 
\cite[Section IV]{Steinberg:05p}, we have that
\begin{align}
\inC(\savbc^{q})=\bigcup_{p(u_1,u_2)\in\Dset^{\,\;\rstarC}} 
\left\{
\begin{array}{lll}
(R_1,R_2) \,:\; & R_2 &\leq    I_q(U_2;Y_2) \,, \\
								& R_1 &\leq   I_q(U_1;Y_1|U_2)  
\end{array}
\right\}\,.
\end{align}
Then, the requirement (\ref{eq:BICachieve1}) is satisfied as well, hence $\encs(u_1,u_2,s)$  and  $\Dset^{\rstarC}$
achieve $\sBIRavc$ and $\sBrICav$. It follows that Condition $\sCond_D$ holds, as $q^*=1$ satisfies the desired property 
 with $\encs(u_1,u_2,s)$ and $\Dset^{\rstarC}$ as described above. 
 

We move to the deterministic code capacity region of the arbitrarily varying BSBC $\savbc$ with causal SI. 
If  $\theta_1=\frac{1}{2}$,  the capacity region is given by $\sBCavc=\sBrCav=\{(0,0)\}$, by (\ref{eq:ex1BrCav}). 
Otherwise, $\theta_0< \frac{1}{2}< \theta_1$, and we now  show  that the condition in Corollary~\ref{coro:BmainDbound} is met.
 Suppose that  $V^{\encs'}_{Y_2|U_2,S}$ is symmetrizable for all $\encs':\Uset_2\times\Sset\rightarrow\Xset$.
That is, for every $\encs'(u_2,s)$,  
there exists $\lambda_{u_2}=J(1|u_2)$ such that
\begin{multline}
(1-\lambda_{u_{b}})W_{Y_2|X,S}(y_2|\encs'(u_{a},0),0)+\lambda_{u_{b}}W_{Y_2|X,S}(y_2|\encs'(u_{a},1),1)    =\\
(1-\lambda_{u_{a}})W_{Y_2|X,S}(y_2|\encs'(u_{b},0),0)+\lambda_{u_{a}}W_{Y_2|X,S}(y_2|\encs'(u_{b},1),1)
\end{multline}
for all  $u_{a},u_{b}\in\Uset_2$, $y_2\in\{0,1\}$. 
If this is the case, then for $\encs'(u_2,s)=u_2+s \mod 2$, taking $u_{a}=0$, $u_{b}=1$, $y_2=1$, we have that
\begin{align}
\label{eq:BSCfair}
(1-\lambda_{1})\cdot(\alpha*\theta_0)+\lambda_{1}\cdot(1-\alpha*\theta_1)= (1-\lambda_{0})\cdot(1-\alpha*\theta_0)+\lambda_{0}\cdot(\alpha*\theta_1) \,.
\end{align}
This is a contradiction. 
 Since $f(\theta)=\alpha*\theta$ is a monotonic increasing function of $\theta$, and since $1-f(\theta)=f(1-\theta)$, we have that  the value of the LHS of (\ref{eq:BSCfair}) is in $[0,\frac{1}{2})$, while the value of the RHS of (\ref{eq:BSCfair}) is in $(\frac{1}{2},1]$.
 Thus, there exists 
$\encs':\Uset_2\times\Sset\rightarrow\Xset$ such that $V^{\encs'}_{Y_2|X,S}$ is non-symmetrizable for $\theta_0< \frac{1}{2}< \theta_1$. 
As Condition $\sCond_D$ holds, we have that $\sBCavc=\sBIRavc=\sBrICav$, due to Corollary~\ref{coro:sBmainDbound}. 
Hence, by (\ref{eq:ex1BrCav}), we have that the capacity region of the arbitrarily varying BSBC $\savbc$ with causal SI is given by  (\ref{eq:Bex1Cavc}). \qed

\section{Analysis of Example~\ref{example:AVBSBC2}}
\label{app:AVBSBC2}
\subsection{Random Parameter BSBC with Correlated Noises}
\label{app:AVBSBC2P1}
Consider the random parameter BSBC $\Brp$ with causal SI. 
By Theorem~\ref{theo:BCrp}, the capacity region of $\Brp$ with degraded message sets with causal SI  is given by $\BCrp=\BICrp$ (see (\ref{eq:BICrp})). Then, to show achievability,
consider the following choice of $p(u_0,u_1)$ and $\encs(u_0,u_1,s)$. Let $U_0$ and $U_1$ be independent random variables, 
\begin{align}
\label{eq:BSBCdistAchieve2}
U_0\sim\text{Bernoulli}\left(\frac{1}{2} \right) \,,\;\text{and}\;\,
U_1\sim\text{Bernoulli}(\beta)  \,,
\end{align}
for $0\leq \beta\leq\frac{1}{2}$, 
 and let 
\begin{align}
\label{eq:BSBCxiAchieve2}
\encs(u_0,u_1,s)=u_0+u_1+s \mod 2\,.
\end{align}
Then,
\begin{align}
 &H_q(Y_1|U_0,U_1)=H_q(S+Z_S)=h(\sdelta) \,,																\nonumber\\ 
 &H_q(Y_1|U_0)=H_q(U_1+S+Z_S)=h(\beta*\sdelta) \,,													\nonumber\\
 &H_q(Y_2|U_0)=H_q(U_1+S+N_S)=h(\beta*\wdelta) \,,								\nonumber\\ 
 &H_q(Y_2)=1 \,, 
\intertext{
where addition is modulo $2$, and $\sdelta,\wdelta$ are given by (\ref{eq:swdelta}). 
Thus,
}
&I_q(U_0;Y_2)=1-h(\beta*\wdelta) \,, \nonumber\\
&I_q(U_1;Y_1|U_0)=h(\beta*\sdelta)-h(\sdelta) \,.
\end{align}
The last inequality on the sum rate in (\ref{eq:BICrp}) is redundant, as shown below.
Since $\theta_0\leq\eps_0\leq\frac{1}{2}$ and $\frac{1}{2}\leq\theta_1\leq\eps_1$, we have that
$\sdelta\leq\wdelta\leq \frac{1}{2}$. Hence,
\begin{align}
& I_q(U_0;Y_2)=1-h(\beta*\wdelta)\leq 1-h(\beta*\sdelta)=I_q(U_0;Y_1) \,,
\end{align}
which implies that $I_q(U_0;Y_2)+I_q(U_1;Y_1|U_0) \leq I_q(U_0,U_1;Y_1 ) $.
 This completes the proof of the direct part.

As for the converse, we need to show that if,
\begin{align}
\label{eq:BrpBSBCconv}
R_1 > h(\beta*\sdelta)-h(\sdelta) \,,
\end{align}
 for some $0\leq \beta\leq \frac{1}{2}$, then it must follows that $R_0\leq 1-h(\beta*\wdelta)$. Indeed, by (\ref{eq:BICrp}) and (\ref{eq:BrpBSBCconv}),
\begin{align}
H_q(Y_1|U_0)>& h(\beta*\sdelta)-h(\sdelta)+H_q(Y_1|U_0,U_1) \nonumber\\
\geq& h(\beta*\sdelta)-h(\sdelta)+\min_{u_0,u_1} H_q(\encs(u_0,u_1,S)+Z_S)\nonumber\\
=& h(\beta*\sdelta)-h(\sdelta)+\min\left( H_q(Z_S),H_q(S+Z_S) \right) \nonumber\\
=& h(\beta*\sdelta)-h(\sdelta)+\min\left( h((1-q)\theta_0+q\theta_1),h(\sdelta) \right) \nonumber\\
=& h(\beta*\sdelta)-h(\sdelta)+h(\sdelta) \nonumber\\
=& h(\beta*\sdelta) \,.
\label{eq:BrpBSBCconv2}
\end{align}
Then, since $\sdelta\leq\wdelta\leq \frac{1}{2}$,  there exists a random variable $L\sim\text{Bernoulli}(\lambda_q)$, with
\begin{align}
  \wdelta=\sdelta*\lambda_q \,,
	\label{eq:BrpBSBCconv3}
\end{align}
for some $0\leq\lambda_q\leq\frac{1}{2}$,
 such that $\tY_2=Y_1+L \mod 2$ is distributed according to 
$\cprob{\tY_2=y_2}{U_0=u_0,U_1=u_1}=\sum_{s\in\Sset} q(s) W_{Y_2|X,S}(y_2|\encs(u_0,u_1,s),s)$.
Thus, 
\begin{align}
H_q(Y_2|U_0)=&H_q(\tY_2|U_0) \stackrel{(a)}{\geq} h\left(  [ h^{-1}(H(Y_1|U_0)) ]*\lambda_q  \right) 
\stackrel{(b)}{\geq} 
h(\beta*\sdelta*\lambda_q)\stackrel{(c)}{=} h(\beta*\wdelta) \,,
\end{align}
where $(a)$ is due to Mrs. Gerber's Lemma \cite{WynerZiv:73p}, and $(b)$-$(c)$ follow from  (\ref{eq:BrpBSBCconv2}) and  (\ref{eq:BrpBSBCconv3}), respectively. \qed

\subsection{Arbitrarily Varying BSBC with Correlated Noises}
\label{app:AVBSBC2P2}
\subsubsection{Without SI}
We begin with the case of an arbitrarily varying BSBC $\Bavcig$ without SI. 
We claim that the single user marginal AVCs $\avc_{1,0}$ and $\avc_{2,0}$  without SI, corresponding to  user 1 and user 2, respectively, have zero capacity.
Denote $q\triangleq q(1)=1-q(0)$. 
Then, observe that the additive noises are distributed according to 
$Z_S\sim\text{Bernoulli}(\eta_q^{(1)})$
and 
$N_S\sim\text{Bernoulli}(\eta_q^{(2)})$
, with
$
\eta_q^{(1)}\triangleq (1-q)\cdot \theta_0+q\cdot \theta_1 
$ and
$
\eta_q^{(2)}\triangleq (1-q)\cdot \eps_0+q\cdot \eps_1 
$, 
for $0\leq q\leq 1$. 
Based on \cite{BBT:60p}, $\opC(\avc_{1,0})\leq\opC^{\rstarC}\hspace{-0.1cm}(\avc_{1,0})= \min_{0\leq q\leq 1} [1-h(\eta_q^{(1)})]$.
Since $\theta_0< \frac{1}{2}\leq \theta_1$, there exists $0\leq q_1\leq 1$
such that $\eta_{q_1}^{(1)}=\frac{1}{2}$, thus $\opC(\avc_{1,0})=0$.
Similarly, $\eps_0< \frac{1}{2}\leq \eps_1$ implies that $\eta_{q_2}^{(2)}=\frac{1}{2}$, for some
$0\leq q_2\leq 1$, thus $\opC(\avc_{2,0})=0$ as well.
The capacity region of the AVBC $\Bavcig$ without SI is then given by  
$\BCavcig=\{(0,0)\}$.

\subsubsection{Causal SI -- Case 1}
Consider the arbitrarily varying BSBC $\avbc$ with causal SI, with
$\theta_0\leq 1-\theta_1\leq \eps_0\leq 1-\eps_1\leq\frac{1}{2}$. 
 By Theorem~\ref{theo:Bmain}, the random code capacity region is bounded by $\BIRavc \subseteq \BrCav\subseteq \BrICav$. We show that the bounds coincide, and are thus tight. 
By (\ref{eq:BcvCI}), (\ref{eq:BICrp}) and (\ref{eq:BIRcompoundP}), we have that 
$
\BrICav=\bigcap_{0\leq q\leq 1} \inC(\Brp) \subseteq 
\inC(\avbc^{q'}) 
$, for every given $0\leq q'\leq 1$. 
 Thus, taking $q'=1$, we have by (\ref{eq:BSBCcorrBCrp}) that
\begin{align}
\label{eq:ex1outB2}
\BrICav
 \subseteq 
\bigcup_{0\leq \beta\leq \frac{1}{2}}
\left\{
\begin{array}{lll}
(R_1,R_2) \,:\; & R_2 &\leq   1-h(\beta*\eps_1) \,, \\
								& R_1 &\leq   h(\beta*\theta_1)-h(\theta_1)
\end{array}
\right\}\,,
\end{align}
where we have used the identity 
$h(\alpha*(1-\delta))=h(\alpha*\delta)$. 

Now, to show that the region above is achievable, we examine the inner bound,
\begin{align}
\BIRavc=
\bigcup_{p(u_0,u_1),\encs(u_0,u_1,s)}\, 
\left\{
\begin{array}{lrl}
(R_0,R_1) \,:\; & R_0 &\leq \min_{0\leq q\leq 1}   I_q(U_0;Y_2) \,, \\
								& R_1 &\leq \min_{0\leq q\leq 1}  I_q(U_1;Y_1|U_0) \,\\
								& R_0+R_1 &\leq \min_{0\leq q\leq 1}  I_q(U_0,U_1;Y_1) 
\end{array}
\right\}\,.
\end{align}
Consider the following choice of $p(u_0,u_1)$ and $\encs(u_0,u_1,s)$. Let $U_0$ and $U_1$ be independent random variables, 
\begin{align}
\label{eq:BSBC2distAchieve}
U_2\sim\text{Bernoulli}\left(\frac{1}{2}\right)  \,,\;\text{and}\;\, U_1\sim\text{Bernoulli}(\beta) \,,
\end{align}
for $0\leq \beta\leq\frac{1}{2}$, 
 and let 
\begin{align}
\label{eq:BSBC2xiAchieve}
\encs(u_0,u_1,s)=u_0+u_1+s \mod 2\,.
\end{align}
Then, as in Subsection~\ref{app:AVBSBC2P1} above, this yields the following inner bound,
\begin{align}
\label{ex1:innerRbsbc2}
\BIRavc \supseteq
\bigcup_{0\leq \beta\leq \frac{1}{2}}\, 
\left\{
\begin{array}{lll}
(R_0,R_1) \,:\; & R_0 &\leq \min_{0\leq q\leq 1}  1-h(\beta*\delta_q^{(2)}) \,, \\
								& R_1 &\leq \min_{0\leq q\leq 1}  h(\beta*\delta_q^{(1)})-h(\delta_q^{(1)})  
\end{array}
\right\}\,.
\end{align}
Note that $\theta_0\leq \delta_q^{(1)}\leq 1-\theta_1 \leq\frac{1}{2}$ and 
$\eps_0\leq \delta_q^{(2)}\leq 1-\eps_1 \leq\frac{1}{2}$. For $0\leq\delta\leq\frac{1}{2}$, the functions 
$g_1(\delta)=1-h(\beta*\delta)$ and $g_2(\delta)= h(\beta*\delta)-h(\delta)$ are monotonic decreasing functions of $\delta$, hence the minima in (\ref{ex1:innerRbsbc2}) are both achieved with $q=1$.
It follows that 
	\begin{align}
	\label{eq:ex1BrCav2}
\BrCav=\BIRavc=\BrICav=
\bigcup_{0\leq \beta\leq 1}\, 
\left\{
\begin{array}{lll}
(R_0,R_1) \,:\; & R_0 &\leq   1-h(\beta*\eps_1) \,, \\
								& R_1 &\leq h(\beta*\theta_1)-h(\theta_1)  
\end{array}
\right\}\,.
\end{align}

It can also be verified that Condition $\sCond$ holds (see Definition~\ref{def:sCondQ} and (\ref{eq:sCondShort})), in agreement with part 2 of Theorem~\ref{theo:Bmain}. 
First, we specify a function $\encs(u_0,u_1,s)$ and a distribution set 
$\Dset^{\rstarC}$ that achieve $\BIRavc$ and $\BrICav$ (see Definition~\ref{def:Bcompoundachieve}).
 Let $\encs(u_0,u_1,s)$ be as in (\ref{eq:BSBC2xiAchieve}), and let $\Dset^{\rstarC}$ be the set of distributions $p(u_0,u_1)$ such that $U_0$ and $U_1$ are independent random variables, distributed according to  (\ref{eq:BSBC2distAchieve}). 
By the derivation above,  the first  requirement in Definition~\ref{def:sCondQ}  is satisfied with
$\Qset=\pSpace(\Sset)$, and by our derivation in 
Subsection~\ref{app:AVBSBC2P1}, 
\begin{align}
\BICrp=\bigcup_{p(u_0,u_1)\in\Dset^{\,\;\rstarC}} 
\left\{
\begin{array}{lrl}
(R_0,R_1) \,:\; & R_0 &\leq    I_q(U_0;Y_2) \,, \\
								& R_1 &\leq   I_q(U_1;Y_1|U_0) \,\\
								& R_0+R_1 &\leq   I_q(U_0,U_1;Y_1) 
\end{array}
\right\}\,.
\end{align}
Then, the second requirement is satisfied as well, hence $\encs(u_0,u_1,s)$  and  $\Dset^{\rstarC}$
achieve $\BIRavc$ and $\BrICav$. It follows that Condition $\sCond$ holds, as $q^*=1$ satisfies the desired property  with $\encs(u_0,u_1,s)$ and $\Dset^{\rstarC}$ as described above. 

We move to the deterministic code capacity region of the arbitrarily varying BSBC $\avbc$ with causal SI.
Consider the following cases.
First, if  $\theta_1=\frac{1}{2}$, then $\eps_1=\frac{1}{2}$ as well, and  the capacity region is given by $\BCavc=\BrCav=\{(0,0)\}$, 
by (\ref{eq:ex1BrCav2}). 
Otherwise, $\theta_0< \frac{1}{2}< \theta_1$.
Then, for the case where $\eps_1=\frac{1}{2}$, 
we show that the random code capacity region, $\BrCav=\{ (R_0,R_1): R_0=0,$ $R_1\leq \opC^{\;\rstarC}(\avc_1)=1-h(\theta_1) \}$ can be achieved by deterministic codes as well.
Based on \cite{CsiszarNarayan:88p,CsiszarKorner:82b}, it suffices to show that there exists a function $\encs:\Uset\times\Sset\rightarrow\Xset$ such that $V^{\encs}_{Y_1|U,S}$ is non-symmetrizable.

Indeed, assume to the contrary that $\theta_0< \frac{1}{2}< \theta_1$, yet  $V^{\encs}_{Y_1|U,S}$ is symmetrizable for all $\encs:\Uset\times\Sset\rightarrow\Xset$. That is, for every $\encs(u,s)$,  
there exists $\sigma_{u}=J(1|u)$ such that
\begin{multline}
(1-\sigma_{u_{b}})W_{Y_1|X,S}(y_1|\encs(u_{a},0),0)+\sigma_{u_{b}}W_{Y_1|X,S}(y_1|\encs(u_{a},1),1)    =\\
(1-\sigma_{u_{a}})W_{Y_1|X,S}(y_1|\encs(u_{b},0),0)+\sigma_{u_{a}}W_{Y_1|X,S}(y_1|\encs(u_{b},1),1)
\end{multline}
for all  $u_{a},u_{b}\in\Uset$, $y_1\in\{0,1\}$. 
If this is the case, then for $\encs(u,s)=u+s \mod 2$, taking $u_{a}=0$, $u_{b}=1$, $y_1=1$, we have that
\begin{align}
\label{eq:BSBC2fair}
(1-\sigma_{1})\theta_0+\sigma_{1}(1-\theta_1)= (1-\sigma_{0})(1-\theta_0)+\sigma_{0}\theta_1 \,.
\end{align}
This is a contradiction, since the value of the LHS of (\ref{eq:BSBC2fair}) is in $[0,\frac{1}{2})$, while the value of the RHS of (\ref{eq:BSBC2fair}) is in $(\frac{1}{2},1]$.
 Hence, $V^{\encs}_{Y_1|U,S}$ is non-symmetrizable, and $\BCavc=\BrCav$.
 
The last case to consider is when $\theta_0\leq\eps_0< \frac{1}{2}<\eps_1\leq \theta_1$.
We now claim that the condition in Corollary~\ref{coro:BmainDbound} is met.
Indeed, the contradiction in (\ref{eq:BSBC2fair}) implies that  $V^{\encs}_{Y_1|U_0,U_1,S}$ is non-symmetrizable with $\encs(u_0,u_1,s)=u_0+u_1+s \mod 2$, given that $\theta_0< \frac{1}{2}< \theta_1$.
Similarly,  $V^{\encs'}_{Y_2|U_0,S}$ is non-symmetrizable with $\encs'(u_0,s)=u_0+s \mod 2$, given that 
$\eps_0< \frac{1}{2}< \eps_1$.
 As Condition $\sCond$ holds, we have that $\BCavc=\BIRavc=\BrICav$, due to Corollary~\ref{coro:BmainDbound}. 
Hence, by (\ref{eq:ex1BrCav2}), we have that the capacity region of the arbitrarily varying BSBC with correlated noises $\avbc$ with causal SI is given by  (\ref{eq:Bex1Cavc2case1}). \qed

\subsubsection{Causal SI -- Case 2}
Consider the arbitrarily varying BSBC $\avbc$ with causal SI, with
$\theta_0\leq 1-\theta_1\leq\;$ $ 1-\eps_1\leq \eps_0\leq \frac{1}{2}$. 
 By Theorem~\ref{theo:Bmain}, 
 the random code capacity region is bounded by $\BIRavc \subseteq \BrCav\subseteq \BrICav$.
Next, we show that the deterministic code capacity region is bounded by 
(\ref{eq:Bex1Cavc2case2inner}) and (\ref{eq:Bex1Cavc2case2outer}). 
 
\begin{proof}[Inner Bound]
Denote
\begin{align}
\label{eq:Bex1Cavc2case2innerA}
\mathsf{A}_{in}\triangleq& \bigcup_{0\leq \beta\leq 1}
\left\{
\begin{array}{lrl}
(R_0,R_1) \,:\; & R_0 &\leq   1-h(\beta*\eps_0) \,, \\
								& R_1 &\leq   h(\beta*\theta_1)-h(\theta_1)
\end{array}
\right\}\,.
\end{align}
We show that $\BIRavc\subseteq\mathsf{A}_{in}$ and $\BIRavc\supseteq\mathsf{A}_{in}$, hence
$\BIRavc=\mathsf{A}_{in}$. As in the proof for case 1 above, consider the set of strategy distributions $\Dset^{\rstarC}$ and  function $\encs(u_0,u_1,s)$ as specified by  (\ref{eq:BSBC2distAchieve}) and (\ref{eq:BSBC2xiAchieve}). Then, this results in the following inner bound,
\begin{align}
\BIRavc\supseteq& \bigcup_{p\in\Dset^{\,\;\rstarC}}
\left\{
\begin{array}{lrl}
(R_0,R_1) \,:\; & R_0 &\leq   \min_{0\leq q\leq 1} I_q(U_0;Y_2) \,, \\
								& R_1 &\leq   \min_{0\leq q\leq 1} I_q(U_1;Y_1|U_0) \,\\
								& R_0+R_1 &\leq \min_{0\leq q\leq 1}  I_q(U_0,U_1;Y_1) 
\end{array}
\right\} \nonumber\\
=& \bigcup_{0\leq\beta\leq\frac{1}{2}}
\left\{
\begin{array}{lll}
(R_0,R_1) \,:\; & R_0 &\leq \min_{0\leq q\leq 1}  1-h(\beta*\delta_q^{(2)}) \,, \\
								& R_1 &\leq \min_{0\leq q\leq 1}  h(\beta*\delta_q^{(1)})-h(\delta_q^{(1)})  
\end{array}
\right\} \nonumber\\
=& \mathsf{A}_{in}
\,,
\label{eq:BSBC2case2b1a}
\end{align}
where the last equality holds since in case 2, we assume that $\theta_0\leq 1-\theta_1\leq \frac{1}{2}$ and 
$ 1-\eps_1\leq \eps_0\leq \frac{1}{2}$.

Now, we upper bound  $\BIRavc$ by
\begin{align}
\BIRavc\subseteq \bigcup_{p(u_0,x)}
\left\{
\begin{array}{lrl}
(R_0,R_1) \,:\; & R_0 &\leq   \min_{s\in\Sset} I_q(U_0;Y_2|S=s) \,, \\
								& R_1 &\leq   \min_{s\in\Sset} I_q(X;Y_1|U_0,S=s) \,\\
								& R_0+R_1 &\leq \min_{s\in\Sset}  I_q(X;Y_1|S=s) 
\end{array}
\right\}\,.
\label{eq:BSBC2case2b1}
\end{align}
We have replaced $U_1$ with $X$ since $(U_0,U_1)\Cbar X\Cbar Y_1$ form a Markov chain. 
 Now, since $X\Cbar (Y_1,S)\Cbar Y_2$ form a Markov chain, the third inequality in (\ref{eq:BSBC2case2b1}) is not necessary.
Furthermore $W_{Y_1|X,S}(y_1|x,1)$ is degraded with respect to $W_{Y_1|X,S}(y_1|x,0)$, whereas
$W_{Y_2|X,S}(y_2|x,0)$ is degraded with respect to $W_{Y_2|X,S}(y_2|x,1)$. Thus,
\begin{align}
\BIRavc\subseteq \bigcup_{p(u_0,x)}
\left\{
\begin{array}{lrl}
(R_0,R_1) \,:\; & R_0 &\leq   I_q(U_0;Y_2|S=0) \,, \\
								& R_1 &\leq   I_q(X;Y_1|U_0,S=1)
\end{array}
\right\}\,.
\label{eq:BSBC2case2b2}
\end{align}
 Observe that the RHS of (\ref{eq:BSBC2case2b2}) is the capacity region of a BSBC without a state, specified by $Y_1=X+Z_1 \mod 2$, $Y_2=X+N_0 \mod 2$ \cite{Bergmans:73p,Gallager:74p}.
 This upper bound can thus be expressed as in the RHS of (\ref{eq:Bex1Cavc2case2innerA}) (see \eg \cite[Example 15.6.5]{CoverThomas:06b}). Hence, $\BIRavc=\mathsf{A}_{in}$, which proves the equality in (\ref{eq:Bex1Cavc2case2inner}).

 To show that the inner bound is achievable by deterministic codes, 
\ie $\BCavc\supseteq\BIRavc$, we consider the following cases.
First, if  $\theta_1=\frac{1}{2}$, then $\eps_0=\frac{1}{2}$ as well, and  $\BIRavc=\{0,0\}$,
by (\ref{eq:Bex1Cavc2case2innerA}).  
Otherwise, $\theta_0< \frac{1}{2}< \theta_1$.
In particular, for $\eps_0=\frac{1}{2}$, we have that 
$
\BIRavc=\{ (R_0,R_1) :$ $ R_0=0 ,$ $ R_1\leq 1-h(\theta_1) \} 
$. 
Then, as shown in the proof of case 1, there exists a function $\encs:\Uset\times\Sset\rightarrow\Xset$ such that $V^{\encs}_{Y_1|U,S}$ is non-symmetrizable.
Thus, based on \cite{CsiszarNarayan:88p,CsiszarKorner:82b},  the deterministic code capacity of user 1 marginal AVC is given by $\opC(\avc_1)
=1-h(\theta_1)$, which implies that $\BIRavc$ is achievable for $\eps_0=\frac{1}{2}$.  

It remains to consider the case where $\theta_0\leq\eps_0< \frac{1}{2}<\eps_1\leq \theta_1$.
By Corollary~\ref{coro:BmainDbound}, in order to show that $\BCavc\supseteq\BIRavc$, it suffices to prove that the capacity region has non-empty interior.
Following the same steps as in the proof of case 1 above, we have that the channels 
$V_{Y_1|U,S}^{\encs}$ and $V_{Y_2|U_0,S}^{\encs'}$ are non-symmetrizable for 
$\encs(u,s)=u+s \mod 2$ and $\encs'(u_0,s)=u_0+s \mod 2$ (see (\ref{eq:BchannelUY1Y2})).
Thus, based on \cite{CsiszarNarayan:88p,CsiszarKorner:82b},  the deterministic code capacity of the marginal AVCs of user 1 and user 2 are positive, which implies that  $\interior{\BCavc}\neq\emptyset$,
hence $\BCavc\supseteq\BIRavc$.
\end{proof}

\begin{proof}[Outer Bound]
Since the deterministic code capacity region is included within the random code capacity region, it follows that $\BCavc\subseteq \BrICav$.
Based on (\ref{eq:BcvCI}), (\ref{eq:BICrp}) and (\ref{eq:BIRcompoundP}), we have that 
$
\BrICav=\bigcap_{0\leq q\leq 1} \inC(\Brp) 
$. Thus, 
\begin{align}
&\BrICav\subseteq
\inC(\avbc^{q=0})\cap \inC(\avbc^{q=1})\nonumber\\
=& \left[ \bigcup_{0\leq\beta\leq \frac{1}{2}} \left\{
\begin{array}{lrl}
(R_0,R_1) \,:\; & R_0 &\leq   1-h(\beta*\eps_0) \,, \\
								& R_1 &\leq   h(\beta*\theta_0)-h(\theta_0)
\end{array}
\right\} \right] 
 \cap
\left[ \bigcup_{0\leq\beta\leq \frac{1}{2}} \left\{
\begin{array}{lrl}
(R_0,R_1) \,:\; & R_0 &\leq   1-h(\beta*\eps_1) \,, \nonumber\\
								& R_1 &\leq   h(\beta*\theta_1)-h(\theta_1)
\end{array}
\right\} \right]
\\
 =& \bigcup_{\substack{\; 0\leq \beta_0\leq 1 \,, \\  0\leq \beta_1\leq 1}}
\left\{
\begin{array}{lrl}
(R_0,R_1) \,:\; & R_0 &\leq   1-h(\beta_0*\eps_0) \,, \\
								& R_0 &\leq   1-h(\beta_1*\eps_1) \,, \\
								& R_1 &\leq   h(\beta_0*\theta_0)-h(\theta_0) \,,\\						
								& R_1 &\leq   h(\beta_1*\theta_1)-h(\theta_1)
\end{array}
\right\}
 \,.
\end{align}
\end{proof}

\end{appendices}

\printbibliography

\end{document}